\documentclass{aa}  

\usepackage{graphicx}
\usepackage{txfonts}
\usepackage{hyperref}
\hypersetup{
    colorlinks=true,
    linkcolor=blue,
    citecolor=blue,
    urlcolor=cyan,
    }
\usepackage{orcidlink}
\usepackage{caption}

\begin{document}

   \title{Chromosphere of the quiet sun}
   \subtitle{I. Shock and current-sheet dynamics and heating}

   \author{Q. Noraz\inst{1,2}\thanks{E-mail: quentin.noraz@astro.uio.no}, M. Carlsson\inst{1,2} \and G. Aulanier\inst{3,1}
          }

   \institute{Rosseland Centre for Solar Physics, University of Oslo, P.O. Box 1029 Blindern, Oslo, NO-0315, Norway\\
             \and
             Institute of Theoretical Astrophysics, University of Oslo, P.O.Box 1029 Blindern, Oslo, NO-0315, Norway.\\
             \and
             Sorbonne Université, Observatoire de Paris – PSL, École Polytechnique, Institut Polytechnique de Paris, CNRS, Laboratoire de Physique des Plasmas (LPP), 4 Place Jussieu, 75005 Paris, France\\
             }

   \date{Received 14 July 2025 / Accepted 30 October 2025}

 
  \abstract
   {The solar chromosphere is a dynamic and crucial interface between the solar interior and its interplanetary environment, regulating how energy is locally deposited into heat and transported into the upper atmospheric layers. Despite significant observational and theoretical progress, the dominant processes responsible for chromospheric heating remain debated, particularly under quiet-Sun (QS) conditions.}
   {We aim to disentangle and quantify the respective roles of shocks and current sheets (CS) in QS chromospheric modeling.}
   {We use a convection-zone-to-corona simulation performed with the radiation-magnetohydrodynamics code \textit{Bifrost}. In order to identify shocks and CS events across space and time, we develop and apply physics-based criteria, allowing us to describe their dynamics and evaluate their contributions to both dissipative (viscous and ohmic) and mechanical (including compressive work) heating.}
   {Shocks are found to dominate the energy deposition in the lower chromosphere ($1 \lesssim z \lesssim 1.5$~Mm), accounting for up to $59\%$ of the mechanical heating rate near $z = 1.2$~Mm. In contrast, CS become the primary contributor in the upper chromosphere ($1.5 \lesssim z \lesssim 2.5$~Mm), as both plasma $\beta$ and Mach number $Ma$ drop. Overall, $66\%$ of the mechanical chromospheric heating is powered by the combined action of shocks and CS, with $13\%$ emerging from regions where shocks and CS overlap, underscoring the pivotal role of dynamic coupling in the chromosphere.}
   {These results support a multi-process view of the chromospheric heating in the QS, dominated by shocks, CS, and non-steep gradient dynamics. In addition to viscous and ohmic dissipation, compressive heating can play a major role locally in the model, particularly in chromospheric shock structures, where it offsets non-reversibly cooling from expansion and radiation, and therefore constitutes a key heating contribution to consider in the energy budget. This study further highlights the need for next-generation observations to resolve the intermittent and small-scale nature of chromospheric dynamics, in order to bring new constraints on the coupling between the different layers of the solar atmosphere.}

   \keywords{magnetohydrodynamics (MHD) --
            shock waves --
            current sheet --
            Sun: atmosphere --
            Sun: chromosphere --
            Methods: Numerical
               }

   \maketitle

\section{Introduction}\label{sec:sect1}
The atmospheric heating of the Sun and solar-type stars is a long-standing mystery in astrophysics. The solar corona indeed reaches million-degree temperatures \citep{grotrianErgebnissePotsdamerExpedition1931,edlenDeutungEmissionslinienIm1943}, although radiative equilibrium would predict a cooler atmosphere beyond the 5772K photosphere \citep{schwarzschildEquilibriumSunsAtmosphere1906,prsaNOMINALVALUESSELECTED2016}. It has thus become of major interest to understand this departure, which starts in the chromosphere.

The chromosphere is the layer lying between the cooler photosphere and hotter corona. Its name reflects the pink shade of its characteristic $H_\alpha$ radiation Balmer's line \citep{deferrerObservationsEclipseSun1809,lockyerSpectroscopicObservationSun1868}, corresponding to the transition from the state n = 3 to n = 2 of the hydrogen atom. Hence, a significant population of these levels is needed to explain observations, which subsequently need the chromosphere temperature to rise above what is predicted by the radiative equilibrium (about 10kK against $\sim 4000$ K expected, see \citealt{cilliePhysicalStateSolar1935,redmanSpectrographicObservationsTotal1942}, \citealt{carlssonNewViewSolar2019} for a recent and extensive review).

Calcium, magnesium, and iron are strong emitters of the Quiet-Sun (QS) chromosphere as well, with an estimated total radiative loss of about $10^7$~erg~cm$^-2$~cm$^-1$ \citep{vernazzaStructureSolarChromosphere1981,andersonModelSolarChromosphere1989}. The photospheric convection is known to contain enough kinetic energy in comparison to the amount needed to sustain atmospheric cooling rates \citep{schwarzschildNoiseArisingSolar1948}. Yet, how this energy reaches and dissipates in the chromosphere remains unclear.

MHD waves \citep{alfvenExistenceElectromagneticHydrodynamicWaves1942} and their dissipation mechanisms (known as AC heating) are being extensively studied in the context of coronal heating (\citealt{arreguiCoronalHeating2024,vandoorsselaereUniturbulenceAlfvenWave2025,cherryDetectionWaveActivity2025,mortonOriginsCoronalAlfvenic2025}). However, the chromosphere comparatively requires higher amounts of energy deposition to balance its radiative losses (10$^6$–10$^7$~erg~cm$^{-2}$~s$^{-1}$ v.s. 10$^4$–10$^6$~erg~cm$^{-2}$~s$^{-1}$ for the solar corona, \citealt{withbroeMassEnergyFlow1977}). Small-scale alfvenic vorticity has been shown to potentially channel a significant portion of the energy flux required (e.g. \citealt{mollVorticesShocksHeating2012,wedemeyer-bohmMagneticTornadoesEnergy2012,shelyagALFVENWAVESSIMULATIONS2013}), however, AC mechanisms alone likely hardly deposit enough of the flux in the QS chromosphere (\citealt{mihalasInternalGravityWaves1981,narainChromosphericCoronalHeating1996}, see also Section 10.3 of \citealt{ulmschneiderHeatingSolarChromosphere2003} and \citealt{srivastavaChromosphericHeatingMHD2021} for a recent review). The solar atmosphere is a high-Reynolds/high-Lundquist numbers environment (\citealt{spitzerPhysicsFullyIonized1956,niFASTMAGNETICRECONNECTION2015,faerderComparativeStudyResistivity2024}). High gradients of either velocity or magnetic field are then required to significantly dissipate, respectively, kinetic or magnetic energy into heat locally. Good candidate mechanisms for this deposition will therefore be shocks \citep{jefferiesMagnetoacousticPortalsBasal2006,wedemeyer-bohmWhatHeatingQuietSun2007,kalkofenSolarChromosphereHeated2007,grantAlfvenWaveDissipation2018,udnaesCharacteristicsAcousticwaveHeating2025} and reconnecting current sheets \citep{hansteenBombsFlaresSurface2017,robinsonIncoherentFieldCoherent2022,robinsonQuietSunFlux2023}.

Shock dissipation is an important process for the deposition of energy from acoustic flux into the chromosphere \citep{biermannZurDeutungChromosphrischen1946,narainChromosphericCoronalHeating1996,wedemeyerNumericalSimulationThreedimensional2004,chaurasiyaPropagationEnergyDissipation2025}, resulting from the surface generation of upwardly propagating acoustic waves \citep{schwarzschildNoiseArisingSolar1948}. In particular, it was shown that Ca H\&K periodic brightenings observed in the QS chromosphere could be successfully reproduced by 1D models of upward propagating shocks \citep{carlssonNonLTERadiatingAcoustic1992,carlssonFormationSolarCalcium1997}. Recent observations report that the deposition of acoustic flux could up to fully balance the radiative losses in the lower chromosphere of some QS areas \citep{abbasvandObservationalStudyChromospheric2020,abbasvandChromosphericHeatingAcoustic2020} and would not be sufficient in higher parts (above 1400 km, see also \citealt{carlssonDoesNonmagneticSolar1995,fossumHighfrequencyAcousticWaves2005, beckEnergyWavesPhotosphere2009, jessMultiwavelengthStudiesMHD2015,molnarHighfrequencyWavePower2021}). This lack of observed acoustic energy has then led to the hypothesis that magnetic-field-related processes likely also play an important role in heating the chromosphere, especially in its upper parts
\citep{carlssonNewViewSolar2019,abbasvandObservationalStudyChromospheric2020}.

Current sheets (CS) in the solar atmosphere generally result either from the slow braiding of magnetic-field lines by horizontal convective motions, or emerging loops interacting with the overlying magnetic canopy \citep{parkerHeatingStellarCorona1986}. They are likely sites of magnetic reconnection (e.g.~\citealt{yamadaMagneticReconnection2010}), releasing energy via DC heating (e.g.~\citealt{ulmschneiderMechanismsChromosphericCoronal2003}). Though initially explored for the corona, the chromosphere’s turbulent dynamics and higher resistivity also favor reconnection \citep{sturrockChromosphericMagneticReconnection1999,niFASTMAGNETICRECONNECTION2015}. However, although several observational signatures of chromospheric reconnection have been reported (see, e.g.,~\citealt{chaeChromosphericMagneticReconnection2007,chaeMagneticReconnectionPhotosphere2012,chittaCompactSolarUV2017,ortizEllermanBombsUV2020}), a detailed quantification of its contribution to the chromospheric heating in the QS is still pending.

Observations now regularly report transport of Poynting flux \citep{jessMultiwavelengthStudiesMHD2015,keysHighresolutionSpectropolarimetricObservations2020} and acoustic flux \citep{abbasvandObservationalStudyChromospheric2020,molnarHighfrequencyWavePower2021} by MHD waves in chromospheric structures, but how much of this flux dissipates in the chromosphere remains unclear \citep{steffensTracingCAGrains1997,molnarHighfrequencyWavePower2021,srivastavaChromosphericHeatingMHD2021}. Chromospheric quantities, such as velocities, magnetic fields and electric currents, are indeed notoriously challenging to infer from observations (see, e.g.,~\citealt{delacruzrodriguezRadiativeDiagnosticsSolar2017,dasilvasantosHeatingSolarChromosphere2022,liTomographySolarPlage2023}), especially when considering under-resolved small-scale dissipation events, such as shocks and CS.

Numerical simulations of the solar atmosphere have recently been developed with some success, in order to reproduce key solar atmospheric features (see, e.g.,~\citealt{rempelRADIATIVEMAGNETOHYDRODYNAMICSIMULATION2009,beeckSimulationsSolarNearsurface2012,martinez-sykoraGenerationSolarSpicules2017,hansteenEllermanBombsUV2019,rempelComprehensiveRadiativeMHD2023}). Especially, it has now been demonstrated that synthetic photospheric spectra show remarkable agreements when compared to high-resolution observations \citep{pereiraHowRealisticAre2013}, but also that the braiding of magnetic field lines by photospheric convection can sustain a million-degree corona via Poynting flux injection \citep{gudiksenInitioApproachSolar2005,hansteenNUMERICALSIMULATIONSCORONAL2015,carlssonPubliclyAvailableSimulation2016,finleyStirringBaseSolar2022}. Details of the energy deposition in the chromosphere of the QS are now investigated and still open for questions \citep{cherryDetectionWaveActivity2025,udnaesCharacteristicsAcousticwaveHeating2025}. In particular, the use of three dimensions (3D) is key to properly model the complex magnetic field topology that impacts the amount of braiding \citep{nobrega-siverioDecipheringSolarCoronal2023}, controls the formation of CS \citep{hansteenEllermanBombsUV2019}, but also the propagation of acoustic waves \citep{jefferiesMagnetoacousticPortalsBasal2006,kontogiannisPowerHaloMagnetic2010,felipeNumericalDeterminationCutoff2020,koyamaSolarChromosphericHeating2024,udnaesCharacteristicsAcousticwaveHeating2025} and their subsequent deposition of energy via shock formation \citep{ulmschneiderValidityAcousticallyHeated2005,kalkofenValidityDynamicalModels2012}.

Leveraging on the pioneering work of \cite{nordlundNumericalSimulationsSolar1982} and \cite{hansteen3DNumericalModels2007}, we use the \textit{Bifrost} code \citep{gudiksenStellarAtmosphereSimulation2011} to investigate: What is the respective role of shocks and reconnecting CS in driving chromospheric heating and dynamics? This paper will aim to answer it by providing numerical constraints for the QS environment. We describe the numerical experiment and its thermal structure in Sect.~\ref{sec:sect2}. Then, we introduce our tracking method for shocks and CS in Sect.~\ref{sec:sect3} before investigating how they drive the chromospheric dynamics. We finally investigate and quantify their contribution to the chromospheric heating in Sect.~\ref{sec:sect4}. We discuss several key points in Sect.~\ref{sec:sect5}, before concluding in Sect.~\ref{sec:sect6}.

\section{Numerical set-up}\label{sec:sect2}
\subsection{Bifrost}\label{sec:Bif}

\begin{figure*}
\sidecaption
    \includegraphics[width=12.5cm]{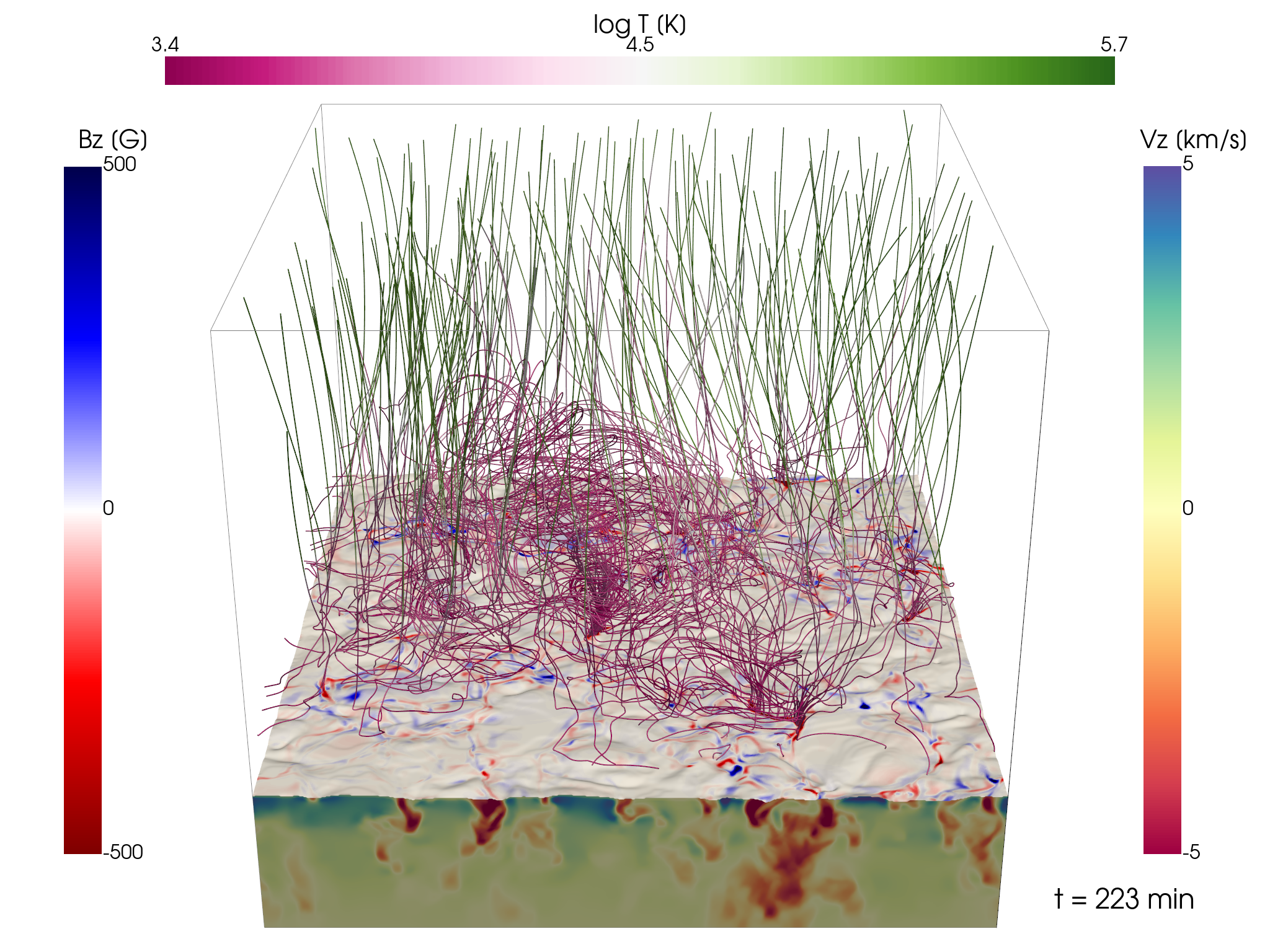}
    \caption{3D visualization of ch012023. The horizontal and corrugated slice represents the $\tau_{500}=1$ surface, coloured with the photospheric vertical magnetic field $B_z$. Vertical convective velocities $v_z$ are illustrated on the side of the upper-convection zone, and magnetic field lines are coloured with temperatures. 15x15 seeds have been uniformly distributed horizontally at 3~Mm to trace magnetic field lines. These lines connect to different structures, and preferentially to magnetic concentration located in intergranular lanes at the photosphere. Magnetic field lines are coloured by temperature, ranging from pink in the temperature minimum region of the chromosphere ($\sim4000$~K), to white in the transition region ($\sim40~000$~K), and finally green when reaching the low corona at the top of our box ($\sim500~000$~K). Note the vortex-like pink feature at the middle of the box, known as a \textit{magnetic tornado}. An animation is available at (To be decided with AA) and covers 12~s, which corresponds to 30 solar minutes.}
    \label{fig:3DglobRend}
\end{figure*}

\textit{Bifrost} is a massively parallel 3D radiative-magnetohydrodynamics (rMHD) code, for which an extensive description can be found in \cite{gudiksenStellarAtmosphereSimulation2011}, and briefly summarized here. It solves explicitly the MHD equations using a sixth-order finite difference spatial scheme along with a fifth-order interpolator on a staggered grid. It also employs explicit third-order time stepping \citep{hymanMethodLinesApproach1979}. The simulation domain is a Cartesian box capturing the low solar atmosphere from the sub-surface convective zone (CZ) to the corona, including the photosphere, chromosphere, and transition region (TR). The equation of state (EOS) for this experiment is tabulated and has been generated with the Uppsala Opacity Package \citep{gustafssonFORTRANProgramCalculating1973,gustafssonGridModelAtmospheres1975}. The simulation implements radiative transfer using multi-group opacities \citep{nordlundNumericalSimulationsSolar1982} with four opacity bins, including scattering \citep{skartlienMultigroupMethodRadiation2000}, and follows a short-characteristics scheme \citep{hayekRadiativeTransferScattering2010}. Hydrogen is treated in local thermodynamic equilibrium (LTE) for this experiment, while the upper atmosphere (chromosphere, TR, and corona) follows the radiative energy budget detailed in \cite{carlssonApproximationsRadiativeCooling2012}. The latter relies on detailed 1D full non-LTE radiative transfer simulations \citep{carlssonNonLTERadiatingAcoustic1992,carlssonDoesNonmagneticSolar1995,carlssonFormationSolarCalcium1997,carlssonDynamicHydrogenIonization2002}. Optically thin radiative losses are prescribed by using tables following atomic data in CHIANTI \citep{dereCHIANTIAtomicDatabase1997,landiCHIANTIAtomicDatabase2006}. Thermal conduction along coronal magnetic field lines, crucial to the upper atmosphere’s energy balance, is modeled using the MURaM code approach \citep{rempelEXTENSIONMURAMRADIATIVE2017}. In this experiment, we do not include any term dealing with ambipolar diffusion or Hall currents (see, e.g.,~\citealt{martinez-sykoraTWODIMENSIONALRADIATIVEMAGNETOHYDRODYNAMIC2012,nobrega-siverioAmbipolarDiffusionBifrost2020} for such aspects, and Sect.~\ref{sec:sect5} for discussion).

The diffusive terms in Bifrost ensure numerical stability by using a global and weak diffusion term alongside a localized hyper-diffusion term. This hyper-diffusion scheme prevents extreme gradients below the grid resolution, such as those formed in collapsing CS or steepening front shocks. By selectively increasing diffusion where needed, the Reynolds and magnetic Reynolds numbers vary across space and time, ensuring a high degree of turbulence on average while preserving numerical stability. This subsequently also regulates magnetic reconnection, preventing numerical instabilities while preserving essential physical processes (see \citealt{faerderComparativeStudyResistivity2023,faerderComparativeStudyResistivity2024} for more details, and Sect.~\ref{sec:sect5} for discussion).

\subsection{Quiet-sun experiment}\label{sec:ch012023}

\begin{figure*}
    \begin{center}
        \includegraphics[width=\linewidth]{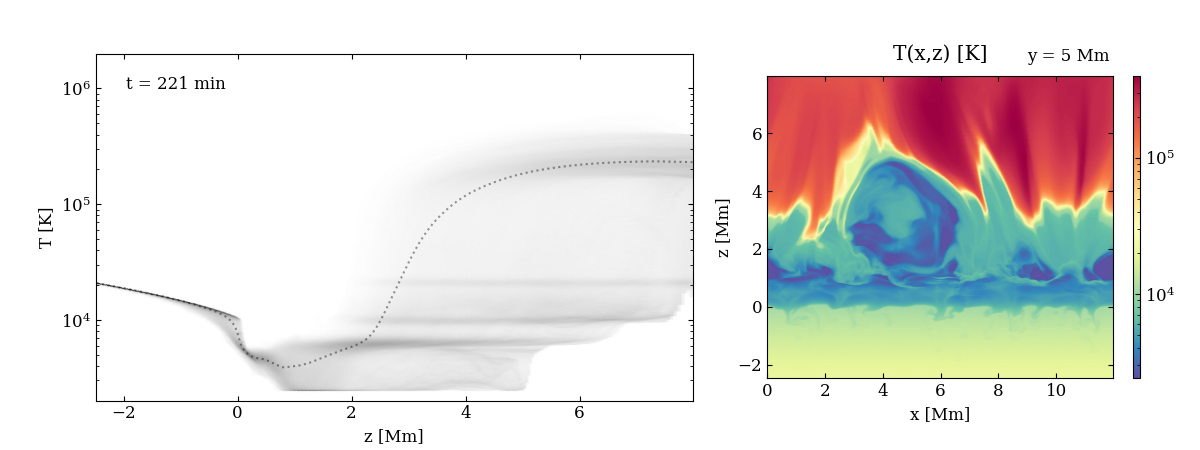}
    \end{center}
    \caption{Temperature distribution. \textit{Left:} Probability density function (PDF) of the temperature as a function of height. The horizontally-averaged mean temperature profile is shown with a dotted line. Note the logarithmic temperature scale. \textit{Right:} Vertical cut in the middle of the domain of the temperature map. Note the 2d cut of the vortex-like structure highlighted in Fig. \ref{fig:3DglobRend} between $x=3$ and 7~Mm, along with spicular structures, notably at $x=8$~Mm, propagating chromospheric temperatures into the transition region. An animation is available at (To be decided with AA) and covers 71~s, which correspond here to 2 solar hours.}\label{fig:tempProfPDFcut}
\end{figure*}

This simulation uses a 512$^3$ grid as the computational domain and \textit{ch012023} as the name, which reflects the Coronal-Hole-like magnetic topology spanning an extent of 12 Mm in both horizontal directions, with a constant horizontal resolution of 23 km. Vertically, it spans from 2.5 Mm below to 8 Mm above the average solar surface ($\tau_{500}=1$) with a non-uniform spacing ranging from 30 km at the bottom of the convective layer, to 14/12 km into the photosphere/chromosphere respectively, and finally increasing to 70.5 km at the coronal boundary. Horizontally periodic boundary conditions (BC) ensure stability, prevent artificial reflections, and assure magnetic connectivity across the domain. The bottom BC is open to inflows, with entropy values adjusted to maintain an effective temperature consistent with the solar one \citep{prsaNOMINALVALUESSELECTED2016}. Over the 2 solar hours analyzed in this paper, we report an averaged value of $T_{\rm eff}=5765$~K, fluctuating from 5745 to 5786~K with a standard deviation of 9~K. At the same time, BC at the top of the box remain open with a characteristics-boundary scheme (\citealt{gudiksenStellarAtmosphereSimulation2011}, see also \citealt{tarrSimulatingPhotosphericCoronal2024}) to allow for free expansion of the plasma. An absorbing layer has been added to the top 7 grid points to improve the limitation of artificial reflections, and hence the chromosphere we aim to model and focus on (see, e.g., the animation of the left panel in Fig.~\ref{fig:shocks}).

This simulation is a continuation of the setup presented in \citep{martinez-sykoraOriginMagneticEnergy2019}, initially configured with a 6~Mm horizontal span and a 5~km pixel size. The model was then successively degraded to 10~km and 20~km resolution. The time origin we consider ($t=0$) is defined from the start of the 20~km-resolution phase. At $t=240$s, the domain was extended via duplication in both horizontal directions and interpolated onto a $512^3$ mesh at the current 23~km horizontal resolution.

The magnetic topology is characterized by an open-field configuration with a 2.5 G horizontally-averaged vertical component, to mimic typical values for QS coronal holes \citep{harveyMagneticMeasurementsCoronal1982,zwaanElementsPatternsSolar1987}. Despite the occurrence of localized kG field concentrations as expected in the QS, the average photospheric magnetic field in our simulation remains weak, with $\langle|Bz|\rangle = 21$~G, $\langle B\rangle = 43$~G, and $B_{\rm rms} = 86$~G when averaged over the $\tau_{500nm}=1$ surface and one solar hour. These values place our setup at the lower end of the parameter space previously investigated in radiative-MHD simulations \citep{rempelNUMERICALSIMULATIONSQUIET2014, khomenkoNumericalSimulationsQuiet2017, przybylskiStructureDynamicsInternetwork2025} and below the levels generally inferred from spectropolarimetric observations \citep{bellotrubioQuietSunMagnetic2019}. Our model thus represents a weakly magnetized quiet-Sun region (see also \citealt{martinez-sykoraOriginMagneticEnergy2019,finleyStirringBaseSolar2022,gosicBifrostModelsQuiet2025}) and constitutes the first case of a future parametric study probing the impact of different magnetic field strengths on chromospheric thermodynamics.

Fig.~\ref{fig:3DglobRend} illustrates a 3D rendering of the simulated solar atmosphere, highlighting the complex interplay between convective motions, magnetic fields, and low-atmospheric dynamics. The side boundaries of the sub-surface CZ reveal hot upflows and cool downflows, which shape the surface granulation patterns. At the photosphere, magnetic field lines emerge preferentially from intergranular lanes, where plasma downflows concentrate the magnetic field into small-scale structures. These regions correspond to magnetic bright points when observed in high-resolution solar images \citep{solankiSmallscaleSolarMagnetic1993,bergerSolarMagneticElements2004,carlssonObservationalManifestationsSolar2004,keysHighresolutionSpectropolarimetricObservations2020}.

Fig.~\ref{fig:3DglobRend} further illustrates the formation of chromospheric swirls (pink-twisted structure), also known as magnetic tornadoes (see \textit{e.g.} \citealt{wedemeyer-bohmMagneticTornadoesEnergy2012, tziotziouVortexMotionsSolar2023}), which originate from these photospheric magnetic concentrations and extend into the upper atmospheric layers. The swirling motions arise from the photosphere, due to the local conservation of angular momentum, as overturning convection spreads and gathers plasma within intergranular lanes, also known as the \textit{bathtub effect} (\citealt{nordlundSolarConvection1985,liuCospatialVelocityMagnetic2019}). At such locations, the plasma $\beta=8\pi P/B^2$ parameter, with $P$ the thermal pressure, has decreased due to the local enhancement of magnetic field amplitude $B$; however, it stays high enough so that the plasma drags the magnetic field along the whirlpool motion and channels the twist along rising magnetic field lines. While these swirls propagate upward, they transition from a plasma-dominated regime (high-$\beta$ plasma) to a magnetically dominated one (low-$\beta$ plasma) in the chromosphere. The plasma then channels along the magnetic structure, which sustains coherent rotational motions up to coronal heights. The open-field configuration in this simulation further facilitates the escape of Alfvénic perturbations and torsional wave energy, contributing to the Poynting flux injection into the upper atmosphere. Such vortex-driven energy transport mechanisms have been extensively studied \citep{finleyStirringBaseSolar2022,breuSwirlsSolarCorona2023,skirvinPoyntingFluxMHD2024} and further linked to heating and acceleration processes in the upper corona and solar wind.

Here, our study focuses on the deposition of energy within the chromosphere in order to understand the resulting heating and temperature structures.

\subsection{Overview of the temperature structures}

In this Section, we illustrate the overall temperature distribution found in the simulation as a function of height. The detailed analysis of the different heating mechanisms and energy balance leading to it will then be further investigated in Sects.~\ref{sec:sect3} and \ref{sec:sect4}.

We illustrate in Fig.~\ref{fig:tempProfPDFcut} temperature structures over the whole domain. The left panel presents the Probability Density Function (PDF) of the temperature as a function of height for $t=84$~min, where we overplot the horizontally averaged mean temperature profile (dotted line). We see a mean temperature structure typical of what is expected in the QS from semi-empirical modelling \citep{vernazzaStructureSolarChromosphere1981}. The photospheric temperature is close to 5765~K, and the cooler chromosphere is separated from the hot corona by the TR. However, the PDF in the background of the Figure clearly illustrates that a horizontal average is far from giving the full picture of temperature structure over the box, especially in the chromosphere and above. 

In the deep CZ, the temperature spread at a given height remains thin, but increases significantly in the subsurface layers (within the last megameter below the photosphere), where hot granular upflows coexist with cool intergranular downflows. A sharp temperature drop is observed around $z=0$, marking the onset of the photosphere, followed by a more constrained temperature range of 3000–5500~K at $z=0.5$~Mm. Moving further upward, the chromosphere emerges, characterized by a mix of both higher and lower temperatures. After an initial decline in the mean temperature profile, an overall steady increase is detected up to $z=2.3$~Mm, driven by various physical processes that will be the focus of this study in the following sections. Between $z=2.3$ and 5 Mm, temperatures span from typical chromospheric values (3 to 10~kK) to TR and coronal temperatures, reaching $T=400$~kK. Above $z=5$~Mm, and up to the top of the domain (8~Mm), the mean temperature stabilizes slightly above $T=200$~kK.

We further note distinct bands of enhanced probability density at temperatures of 6~kK, 10~kK, and 22~kK. These correspond to the ionization of \ion{H}{i}, \ion{He}{i}, and \ion{He}{ii}, respectively, which are treated under local thermodynamic equilibrium (LTE) in this simulation. Such processes act like thermostats, regulating the temperature to the corresponding ionization energy, hence increasing the density of the PDF accordingly (see \citealt{goldingNONEQUILIBRIUMHELIUMIONIZATION2016} for a detailed discussion). Additionally, a lower temperature limit of 2400~K is enforced by an artificial heating term that activates when temperatures drop below 2500~K. This constraint is necessary to prevent unphysically low temperatures in regions experiencing rapid expansion, such as those influenced by magnetic loop emergence (see \citealt{leenaartsMinimumTemperatureQuiet2011} for a detailed discussion).

Although PDFs provide a statistical perspective on variations of the temperature range with height, they do not easily reveal the underlying spatial structures. In the right panel of Fig.~\ref{fig:tempProfPDFcut}, we present a vertical cut of the temperature field to illustrate them. At the base of the photosphere ($z\simeq 0$), we observe the granular structures, where cool downdrafts can be distinguished as sinking blue tongues. Moving higher into the chromosphere (1-2~Mm), cooler regions (blue, $T\sim 3000$~K) are also visible. These pockets primarily result from the rarefaction wave following upward-propagating shock fronts, as well as from the adiabatic expansion associated with emerging magnetic structures \citep{leenaartsMinimumTemperatureQuiet2011}, as can be further seen in the associated animation (To be agreed with AA). A noteworthy structure is the broad circular feature spanning from $x=3$ to $x=7$~Mm and extending up to $z=5$~Mm. This structure corresponds to a vertical cross-section of a magnetic tornado, further illustrated in Fig.~\ref{fig:3DglobRend}.

Indeed, the right panel illustrates a variety of dynamical structures that transport cool chromospheric plasma up to the TR, leading to a wide range of temperatures between 3 and 5~Mm. Nonetheless, when considering individual vertical columns, the transition from upper chromospheric temperatures (green, $T\sim10$~kK) to coronal temperatures (orange/red, $T>100$~kK) remains sharp over height: see the narrow and corrugated yellow layer ($T\sim30$~kK plasma).

In the next Sect.~\ref{sec:sect3}, we focus on identifying features that are good candidates to actually sustain hot-chromospheric temperatures, and further quantify their respective contributions to the overall chromospheric heating in Sect.~\ref{sec:sect4}.

\section{High-gradient processes}\label{sec:sect3}
In the high-Reynolds/ high-Lundquist number environment of the solar atmosphere, strong gradients are necessary to efficiently convert kinetic and magnetic energy into heat \citep{ulmschneiderHeatingSolarChromosphere2003}. In this Section, we aim to locate such small-scale processes in the model and further discuss their overall dynamics in the system.

\begin{figure*}
    \begin{center}
        \includegraphics[width=0.49\linewidth]{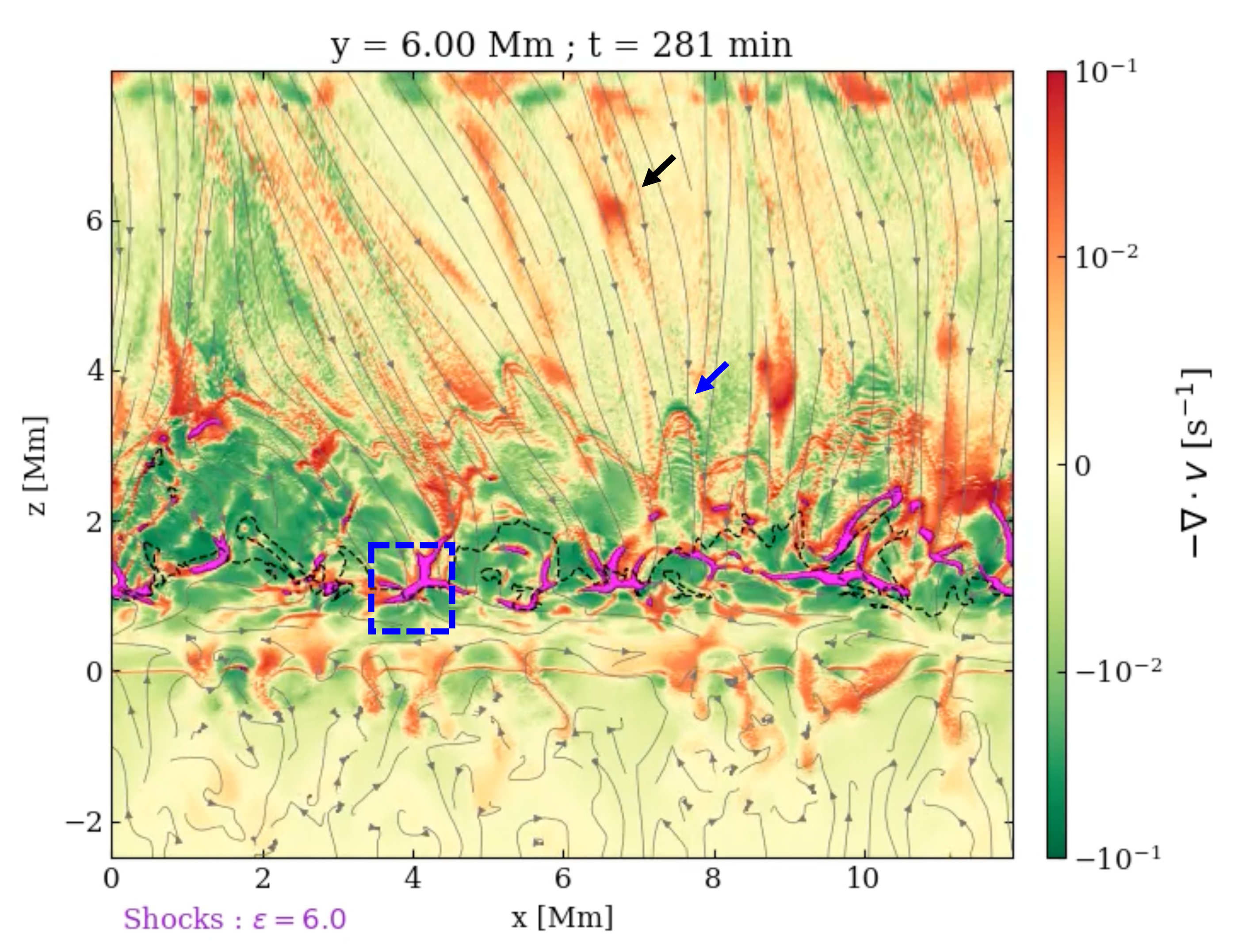}
        \includegraphics[width=0.49\linewidth]{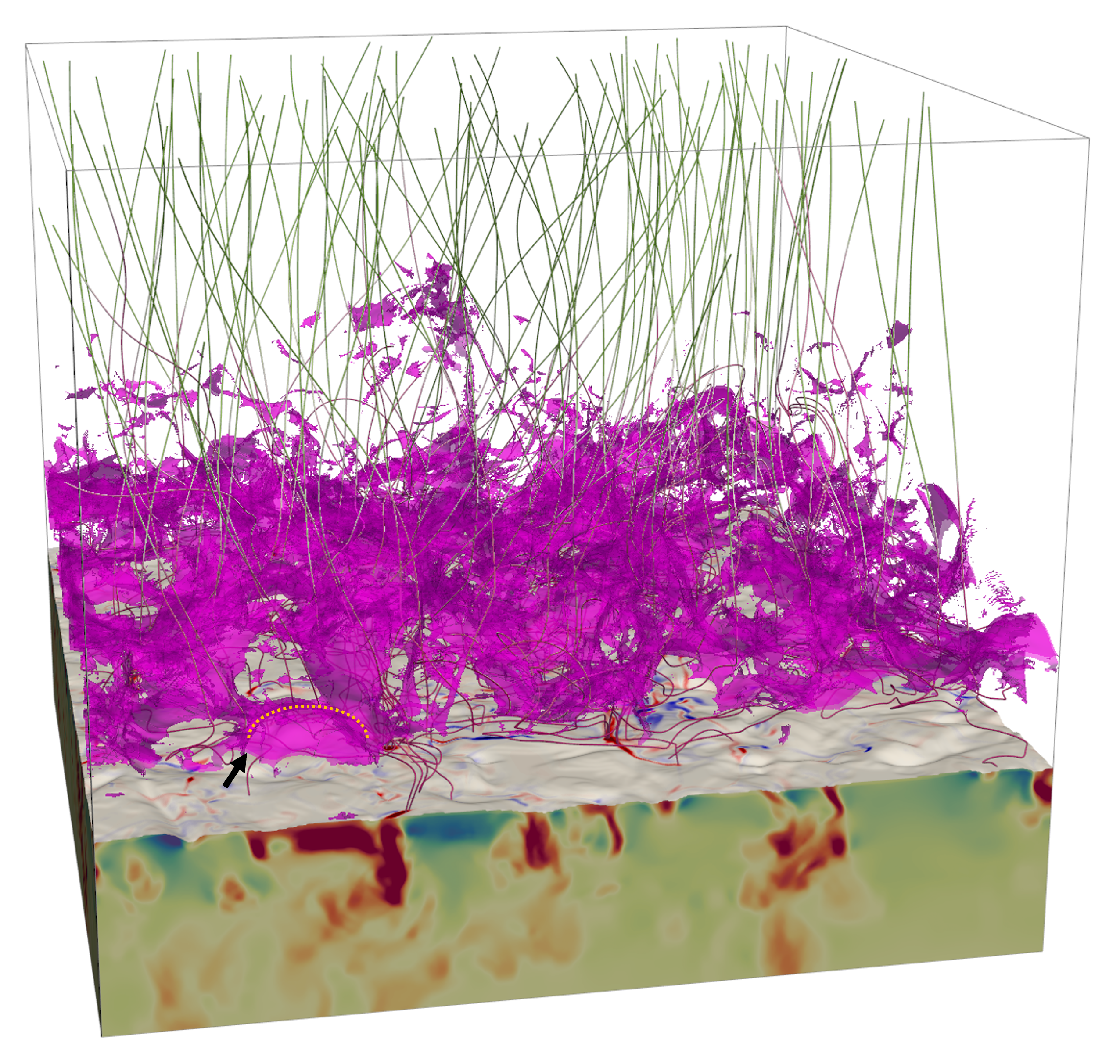}
        \includegraphics[width=\linewidth]{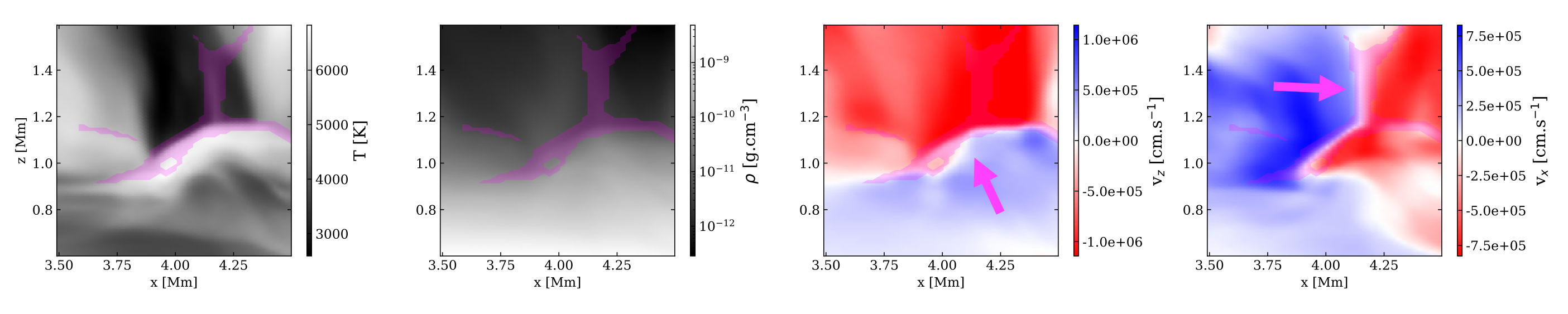}
    \end{center}
    \caption{Shock locations. \textit{Top-left:} The compression frequency $-\mathbf{\nabla}\cdot\mathbf{v}$ taken at Y = 6 Mm. Red and green areas correspond to compression and expansion, respectively, and the magnetic polarity is illustrated along magnetic field lines (grey) with arrows. A zoom of the blue-dashed square is proposed in the bottom panels to highlight shock features (purple). Dark and blue arrows are used to highlight particular compression locations due to wave propagation and rising over-density, respectively. The $\beta=1$ surface is illustrated with a dashed black line. An animation is available at (To be decided with AA) and covers 13~s, which corresponds to 21 solar minutes. \textit{Top-right:} 3D rendering of the shock fronts in the simulation domain at t = 274 min. Similar to Fig.~\ref{fig:3DglobRend}, but we do not colour magnetic field lines for the sake of clarity. Note that we use an orange dashed line along with a dark arrow to highlight the dome-like shape of a shock's front. The locations of the magneto-acoustic shocks are highlighted in magenta for both panels, identified with large values of compression ($-\mathbf{\nabla}\cdot\mathbf{v}>c_s/(6.ds)$, see Eq.~\ref{eq:cs_crit}). An animation is available at (To be decided with AA) and covers 11~s, which corresponds to 18 solar minutes. \textit{Bottom:} A zoom on shock features, showing from left to right the temperature, density, vertical and horizontal velocity along x, respectively. We note that the purple contours overlay the strong gradients of the different quantities. Indicative purple arrows are shown to illustrate the shock motions.
    }\label{fig:shocks}
\end{figure*}

\subsection{Detection and evolution of shocks}\label{sec:shocks}

A first strong gradient of interest is the velocity one, which leads us to track shock fronts. Magneto-sonic waves rising through the chromosphere can indeed produce shocks \citep{schwarzschildNoiseArisingSolar1948,carlssonNonLTERadiatingAcoustic1992}. To locate them in the simulation, we use the following sonic-compression criteria (see also \citealt{wangSimulationAlfvenWave2020,finleyStirringBaseSolar2022}) 
\begin{equation}
    -\mathbf{\nabla}\cdot\mathbf{v}>\frac{c_s}{\epsilon.ds},
    \label{eq:cs_crit}
\end{equation}
where $c_s$ is the sound speed computed from the tabulated EOS we use, such that
\begin{equation}
    c_s = \left(\dfrac{\partial P}{\partial\rho}\right)_{e_{\rm int}}+\dfrac{e_{\rm int}+P}{\rho}\left(\dfrac{\partial P}{\partial e_{\rm int}}\right)_\rho,
    \label{eq:cs_eos}
\end{equation}
with the mass density $\rho$, and the internal energy per unit of volume $e_{\rm int}$. $-\mathbf{\nabla}\cdot\mathbf{v}$ quantifies the local-compression frequency of the plasma and this sonic-compression criteria (Eq.~\ref{eq:cs_crit}) aim at finding locations where the timescale of this compression is faster than the propagation time of a sound wave over the distance $\epsilon.ds$, with $ds={\rm max}(dx,dy,dz)$ and $\epsilon$ being a dimensionless parameter representing a number of grid points. $\epsilon$ is calibrated for each experiment to distinguish linear-wave propagation from acoustic shocks. This selection can be challenging when it comes to solely differentiating high-amplitude waves from weak shock fronts. Here, we set $\epsilon=6$ in order to be very conservative. We may miss a minority of shocks, but ensure that we do not select what one could consider as high-amplitude linear waves (see  Appendix~\ref{sec:appA}).

We illustrate the location of shocks in Fig.~\ref{fig:shocks}. The left panel shows a vertical cut of the compression frequency $-\mathbf{\nabla}\cdot\mathbf{v}$ at $y=6$~Mm, where we highlight the presence of shocks with purple contours. The locations of compression and expansion are illustrated in red and green, respectively. As expected from Eq.~$\ref{eq:cs_crit}$, the strongest shocks correspond to areas of high negative divergence, indicating intense local compression. Notably, we also highlight the presence of weaker-amplitude compression/expansion alternating patterns (indicated by dark and blue arrows), which do not meet the criteria for classification as shocks.

Focusing on the purple features, we see these contours indeed overlap with steep gradients of different quantities (bottom panels). Especially, they mark substantial local temperature enhancements (left panel), as expected from shock propagation (see Appendix~\ref{sec:appC}). This also coincides with density increases due to advection and compression (second panel), as the one resulting from the upward-propagating shock illustrated with a purple arrow in the third panel.

An animation accompanying the left panel of Figure~\ref{fig:shocks} illustrates the predominantly upward propagation of these shocks, driven by granular convection in the photosphere. Below the $\beta=1$ surface (dashed black line), shocks propagation is preferentially upwards and dominates over the magnetic field. Between the photosphere and the $\beta=1$ surface, we note that magnetic field lines are notably twisted and disturbed, due to the convective-overturn and shocks propagation, as will be further illustrated in Sect.~\ref{sec:interplay}. Above the $\beta=1$ surface, most shocks continue their upward trajectory by following the field lines, as expected for slow-mode perturbations in a low-$\beta$ regime. Nonetheless, we note some shock fronts that exhibit horizontal deflections perpendicular to magnetic field lines, suggesting the propagation of fast-mode perturbations in the $\beta<1$ regime (see, e.g., the bottom-right panel).

Such complexity of shock front propagation in our model is further highlighted by its 3D nature. 3D effects are crucial for modeling acoustic wave and shock propagation accurately, as well as their role in acoustic heating \citep{kalkofenValidityDynamicalModels2012}. Previous studies have emphasized that one- or two-dimensional models likely overestimate acoustic heating by artificially increasing the probability of shock front merging, but also oversimplifying wave front shapes, such as plane parallel \citep{ulmschneiderValidityAcousticallyHeated2005}. In the right panel of Figure~\ref{fig:shocks}, we highlight a typical shock front with a yellow dotted line and a black arrow. This exhibits a characteristic dome-like structure, as a direct consequence of 3D acoustic wave propagation. An animation of this panel further illustrates the complex evolution when interacting with other shocks and their environment.

\begin{figure*}
    \begin{center}
        \includegraphics[width=0.49\linewidth]{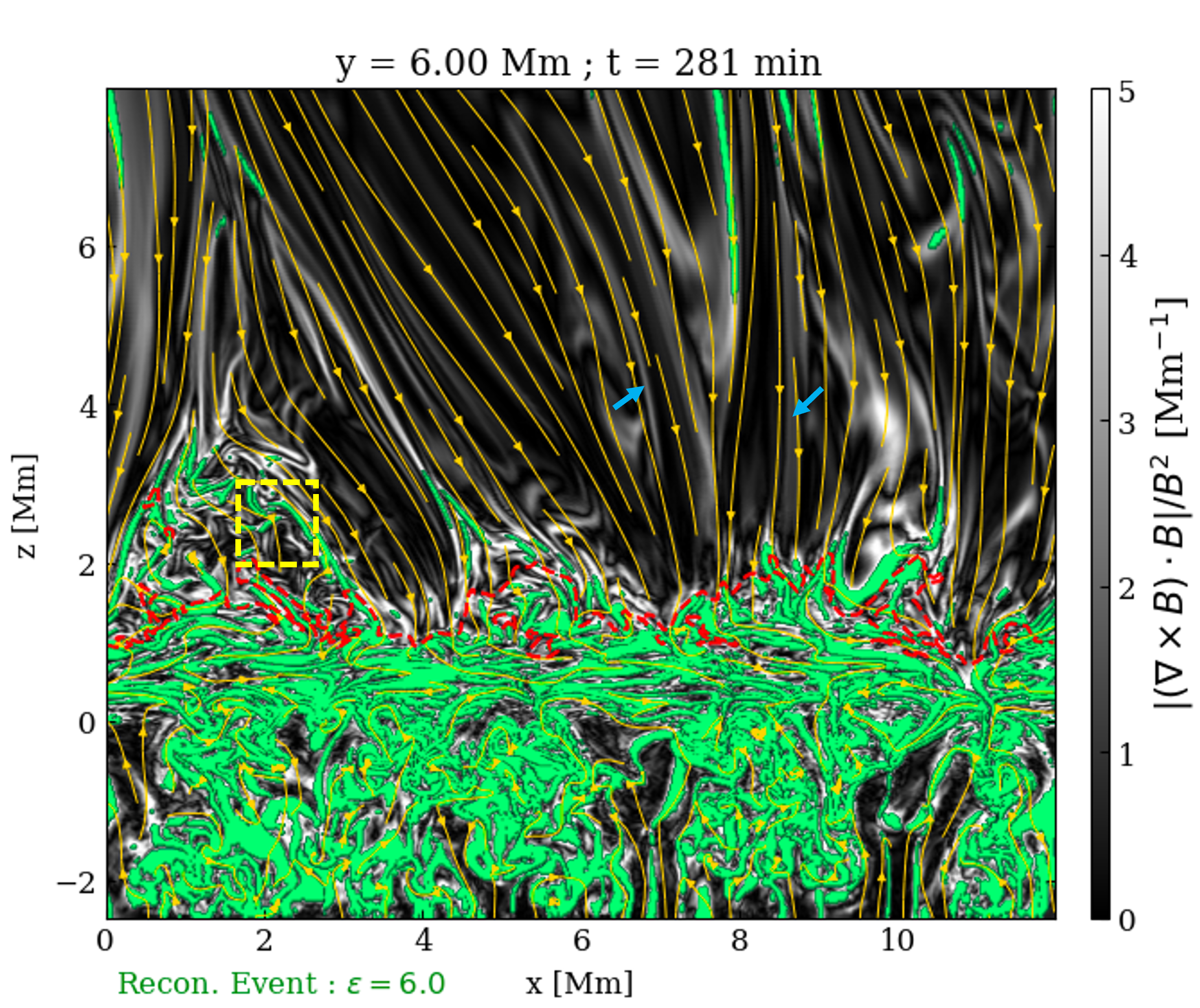}
        \includegraphics[width=0.49\linewidth]{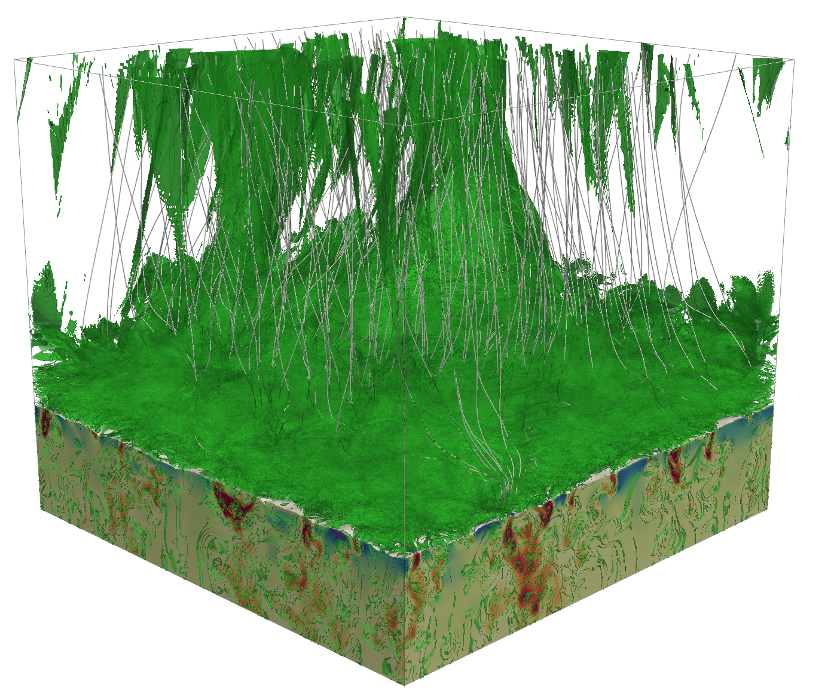}
        \includegraphics[width=\linewidth]{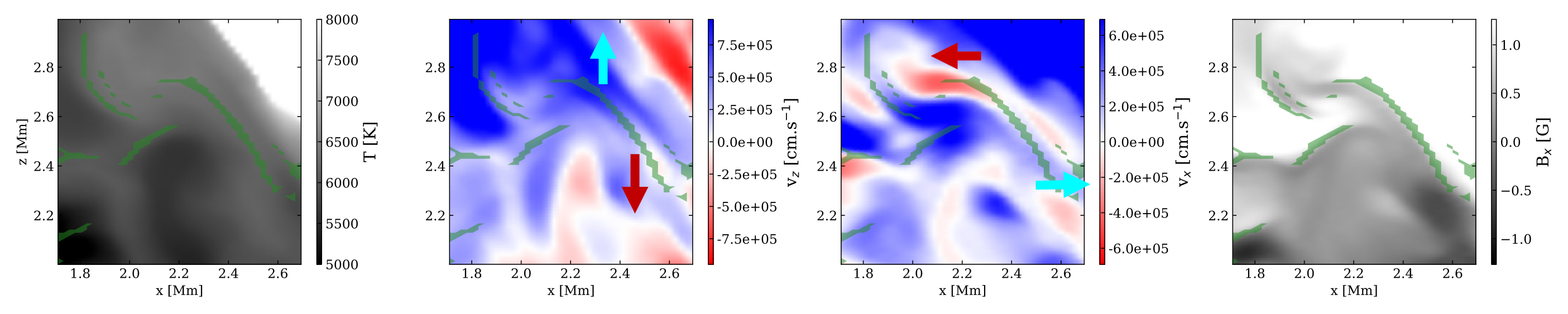}
    \end{center}
    \caption{CS location. \textit{Top-left:} The normalized-parallel current $|\mathbf{\nabla}\times B\cdot B|/B^2$ taken at $y=6$~Mm. Note that this quantity has the dimension of a spatial frequency, with white areas corresponding to high amplitudes. The magnetic polarity is illustrated along magnetic field lines (yellow) with arrows. A zoom of the yellow-dashed square is proposed in the bottom panels to highlight a CS feature. The $\beta=1$ surface is illustrated with a red dashed line. Blue arrows are used to highlight particular locations of weaker and broader current layers, not labelled as CS following Eq.~\ref{eq:alpha_crit}, and likely due to wave propagation, magnetic field braiding, and phase mixing. The $\beta=1$ surface is illustrated with a dashed red line. An animation is available at (To be decided with AA) and covers 12~s, which corresponds to 20 solar minutes. \textit{Top-right:} 3D rendering of CS in the simulation domain at t = 224 min. The locations of CS are highlighted in green for both panels, identified with large values of the normalized-parallel current ($|(\mathbf{\nabla}\times B)\cdot B|/B^2>1/(6.ds)$, see Eq.~\ref{eq:alpha_crit}). \textit{Bottom:} A zoom on CS features, showing from left to right the temperature, vertical velocity, horizontal velocity, and magnetic field component along x, respectively. We note that the green contours overlay the strong gradients in $B_x$ and bipolar velocity patterns. Indicative arrows are shown in cyan and dark red to illustrate these motions.
    }\label{fig:currentS_detection}
\end{figure*}

Finally, the left panel shows the presence of compression structures (red regions) that do not satisfy the sonic compression criterion (Eq.~\ref{eq:cs_crit}). We highlight two examples of such structures, marked with blue and black arrows, near the TR and corona, respectively. From the upper chromosphere to the TR, the local sound speed increases sharply due to the steep temperature gradient. As shocks rise up in the TR, the local sound-speed subsequently overpasses the shock speed, causing shock fronts to broaden and transition back into linear wave propagation (black arrow), with the remaining acoustic and Poynting flux transported upward into the corona. The second compression feature, highlighted by the blue arrow, results from a localized overdensity. As the shock propagates through the chromosphere, the associated pressure gradient accelerates plasma upward. However, as the front broadens upon entering the TR, the pressure gradient weakens, allowing the perturbation (black) to detach from the rising plasma (blue). The plasma (blue arrow) continues to rise due to its residual momentum but also decelerates due to the gravitational force, while the linear wave component continues its ascent at the local sound speed (black arrow). Both structures follow magnetic field lines, as expected for acoustic perturbations in a low-$\beta$ regime. This mass-loading mechanism is consistent with the formation of type-I spicules via acoustic shock propagation (blue arrow, \citealt{hansteenDynamicFibrilsAre2006,depontieuTaleTwoSpicules2007,srivastavaRelationSolarSpicules2025}). This upward propagation of shocks, transitioning back into linear waves, injects a substantial flux of slow-mode MHD waves into the corona. Their exact role in heating the upper parts of the solar atmosphere is still investigated \citep{vandoorsselaereCoronalHeatingMHD2020,cherryDetectionWaveActivity2025}. 

\subsection{Detection and evolution of current sheets}\label{sec:currentS}

The second quantity of interest when looking at high gradients in magnetized plasmas is the magnetic field. This naturally leads us to track current sheets (CS), where sharp variations occur. These structures play a crucial role in magnetic reconnection, energy dissipation, and plasma dynamics. To locate them in the simulation, we adopt the following normalized-parallel-current criterion

\begin{equation} 
    \frac{|\mathbf{\nabla} \times \mathbf{B} \cdot \mathbf{B}|}{\mathbf{B}^2} > \frac{1}{\epsilon\cdot ds}, 
    \label{eq:alpha_crit} 
\end{equation}

where $\mathbf{B}$ is the magnetic field. Similarly to Eq.~\ref{eq:cs_crit}, we set here $\epsilon=6$ to ensure that we distinguish CS from broader current layers with weaker magnetic gradients. This diagnostic is a restrictive form of the commonly employed force-free parameter $\alpha=J/B$ parameter commonly used in the community (see e.g \citealt{aulanierCurrentSheetFormation2005,inoueMHDModelingSuccessive2021,robinsonQuietSunFlux2023}).

To complement the identification of CS, we indeed focus on regions where magnetic reconnection occurs, as CS are necessary sites for such processes (see, e.g., \citealt{wangNumericalSimulationOscillatory2025} and references therein). In the framework of the general theory of reconnection, these sites are characterized by a significant parallel electric field $E_{\parallel}$ \citep{hesseTheoreticalFoundationGeneral1988,schindlerGeneralMagneticReconnection1988}. In our model, such locations correspond to strong values of $|\mathbf{\nabla} \times \mathbf{B} \cdot \mathbf{B}|/\mathbf{B}^2$. By thresholding this quantity, we isolate the dynamically relevant core regions of CS, where reconnection occurs.

We illustrate the location of CS in Fig.~\ref{fig:currentS_detection}. The left panel shows a vertical cut of the normalized-parallel current $|\mathbf{\nabla}\times \mathbf{B}\cdot \mathbf{B}|/\mathbf{B}^2$ at $y=6$~Mm, where we highlight the presence of CS with green contours. The background map represents the strength of the normalized parallel current, with white and black colours indicating strong and weak amplitudes, respectively. The red-dashed line marks the $\beta =1$ surface, delineating regions where plasma and magnetic pressures have similar amplitudes. Magnetic field lines are shown in yellow, showing how their topology relates to the detected CS. As expected from Eq.~$\ref{eq:alpha_crit}$, CS occur in regions where we note variations of the magnetic field orientation.

CS fill a significant portion of the volume below the $\beta=1$ surface (red dashed line), indicating that these structures primarily emerge in a high-$\beta$ environment where plasma pressure dominates over the magnetic one. We especially note a strong correlation of CS location with convective down-flows, as illustrated by the animation attached to Fig.~\ref{fig:currentS_detection}. As convective cells overturn in the photosphere (between $z=0$ and 300~km), they concentrate CS in regions of converging plasma flows, where shearing effects are most intense. Additionally, the overturning motion of convection also leads to the formation and accumulation of horizontal CS in the upper photosphere. This supports the idea that photospheric motions play a key role in structuring the small-scale magnetic field \citep{gallowayConvectionMagneticFields1981,stenfloDistributionFunctionsMagnetic2010,borreroInferringMagneticField2013,bellotrubioQuietSunMagnetic2019} and shaping the formation of CS in the QS \citep{isobeConvectiondrivenEmergenceSmallScale2008,dahlburgTurbulentCoronalHeating2012,chittaFleetingSmallscaleSurface2023}.

Conversely, fewer CS are observed above the $\beta=1$ surface. In this regime, the magnetic field dominates the dynamics, and more especially the magnetic tension force counteracts the shearing imposed by plasma flows, making the formation of sharp gradients and strong currents more difficult. The magnetic field is organized as larger structures rather than fragmented CS. Although fewer CS are present in the low-$\beta$ regime, it is important to recall here that the magnetic energy exceeds the internal energy in such environment ($\beta\sim e_{\rm int}/e_{\rm mag}<1$), which means that the transfer term from the former to the latter, i.e the ohmic heating, is able to heat in these CS significantly. As we will explore in the next Sect.~\ref{sec:location}, this has direct implications for chromospheric heating and temperature structures.

Focusing on the region highlighted by the dashed yellow box (bottom panels), we confirm that the green contours overlay magnetic discontinuities (right panel) and coincide with local temperature enhancements (left panel). We also identify bipolar velocity structures consistent with reconnection outflows, as expected from CS in a low-$\beta$ environment, which further illustrate the focus of our detection on the core dynamical part of CS.

The right panel of Fig.~\ref{fig:currentS_detection} presents a 3D rendering of the simulation domain, where the green structures correspond to CS detected via the same criterion Eq.~$\ref{eq:alpha_crit}$. The white surface represents the $\tau_{\rm 500 nm} = 1$ layer, which outlines the base of the photosphere, but is largely obscured by the high density of detected CS. This visualization emphasizes the complex and layered structure of CS, the dynamics of which is further investigated in the next Section.

It is finally interesting to note the presence of weaker and broader current layers, aligned along vertical magnetic field lines in the upper atmosphere in the left panel. We indicate two of them using blue arrows. These current layers exhibit larger characteristic length scales $B^2/|\mathbf{\nabla}\times B\cdot B|$, suggesting that they originate from either MHD-wave propagation illustrated in Fig.~\ref{fig:shocks}, or magnetic braiding induced via footpoint photospheric motions \citep{gudiksenInitioApproachSolar2005a,hansteenNUMERICALSIMULATIONSCORONAL2015} and subsequent phase-mixing \citep{howsonPhaseMixingWave2020}.

\begin{figure}
    \begin{center}
        \includegraphics[width=\linewidth]{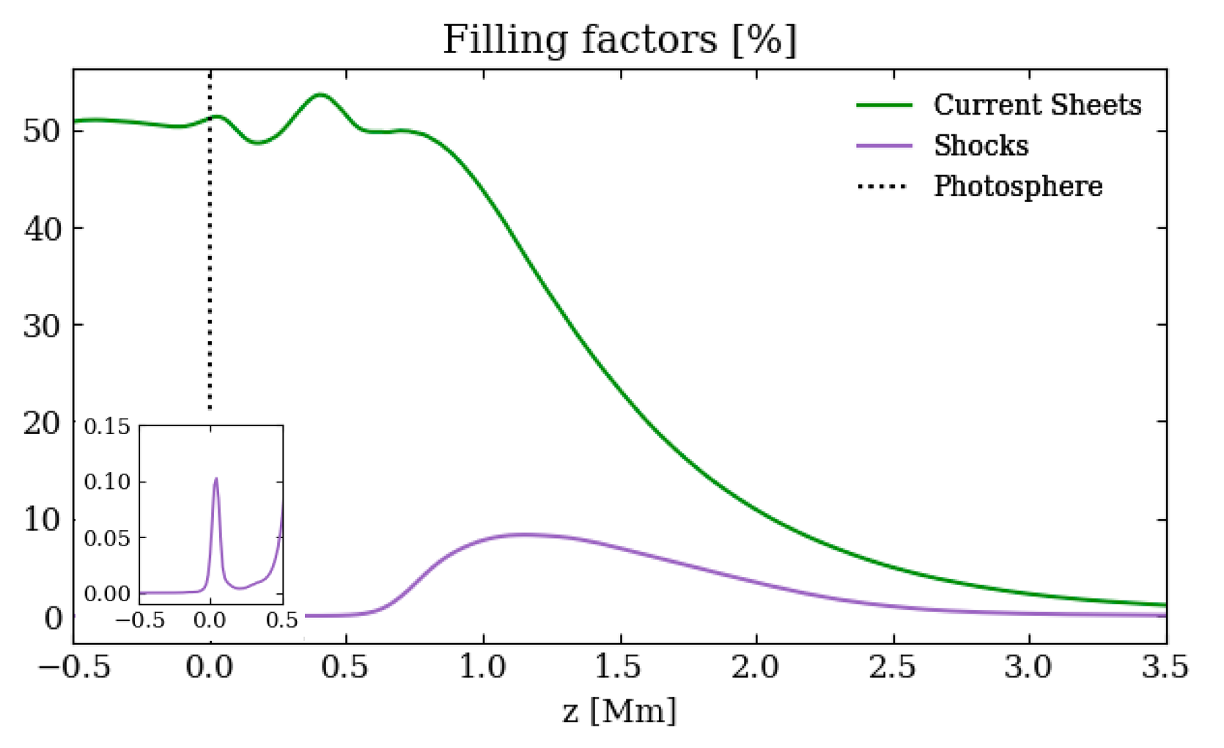}
    \end{center}
    \caption{Fractional number of grid cells labelled as shocks (purple line) or CS (green line), as a function of height. An indicative dotted vertical line shows the position $z=0$~Mm (bottom of the photosphere), and a zoom is proposed to catch the small peak of shocks filling factor at this location. These curves are averaged over one hour in time and horizontally in space.}\label{fig:fill_fact}
\end{figure}

\begin{figure*}
    \begin{center}
        \includegraphics[width=\linewidth]{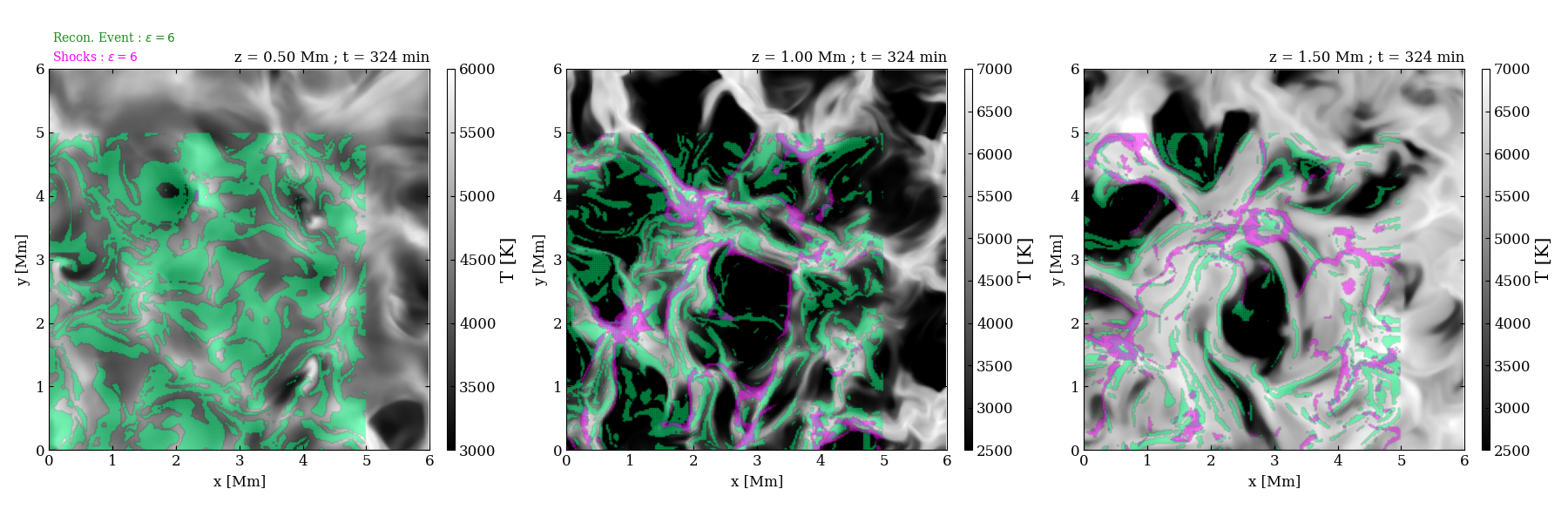}
    \end{center}
    \caption{Interplay of shocks (purple) and CS (green) with temperature structures in the chromosphere (greyscale colorbar). A zoom on $6\times 6$~Mm$^2$ is shown to emphasise the chromospheric structures. Shocks and CS overlay are only considered on a $5\times 5$~Mm$^2$ portion of the panels, to clearly illustrate the overlap between high-gradient processes and temperature structures. An animation is available at (To be decided with AA) and covers 29~s, which corresponds to 48 solar minutes.
    }\label{fig:HorizLoc}
\end{figure*}

\subsection{Filling factor}\label{sec:location}

Fig.~\ref{fig:fill_fact} shows the filling factor of shocks (purple) and CS (green) as a function of height, averaged over 1 hour of solar time and spatially over the horizontal plane. This illustrates the fractional number of grid cells (see Sect.~\ref{sec:sect5} for discussion) that satisfy the criteria of Eqs.~\ref{eq:cs_crit} and \ref{eq:alpha_crit}, respectively, as a function of height.

We note that shocks appear slightly after $z=500$~km. This is consistent with the idea that the chromosphere starts once dissipations become strong enough to break radiative equilibrium. Indeed, acoustic waves are first continuously generated by granular motions at the solar surface and transport kinetic energy into the upper layers via the acoustic wave-energy flux, $F_{\rm acc}$. Given that the mean upward granular velocity in our model ($\langle v_{v_z>0} \rangle_{z=0}\simeq 3.5$~km~s$^{-1}$) is lower than the local sound speed ($\langle c_s \rangle_{z=0}=8.3$~km~s$^{-1}$), acoustic waves are generated with a relatively small amplitude and propagate freely through the photosphere without significant dissipation. However, as the density and temperature decrease with height, the sound speed also decreases. These combined effects lead to a sharp increase in the wave velocity due to the conservation of $F_{\rm acc}=\rho v^2 c_s/2$, until the wave shocks and dissipate enough. The temperature then starts to increase with height, which defines the temperature minimum and the beginning of the chromosphere. We refer the interested reader to Chapters 5 of \cite{mihalasFoundationsRadiationHydrodynamics1984} and \cite{priestMagnetohydrodynamicsSun2014} for more details about shock formation, structure, and evolution.

The filling factor of shocks reaches a maximum of $\sim 9\%$ at $z=1.2$~Mm. This height is consistent with earlier 1D simulations of acoustic shocks in the solar atmosphere \citep{carlssonNonLTERadiatingAcoustic1992,carlssonFormationSolarCalcium1997}. Above $1.2$~Mm, the filling factor of shocks decreases significantly. Its decrease in the upper chromosphere is primarily due to a decrease of the acoustic flux transported, as part has already been deposited lower down. A cause of this occurrence decrease can also be deflections when entering the low-$\beta$ regime \citep{bogdanWavesMagnetizedSolar2003,enerhaugIdentifyingMagnetohydrodynamicWave2025}. The occurrence decrease in the TR can also be linked to the steep rise in temperature, which increases the local sound speed. When the sound speed exceeds the velocity of the shock front, the shock broadens and transitions back into a propagating linear wave (see Fig.~\ref{fig:shocks} and related animations).

A secondary peak is visible near $z=0$~Mm, reaching about $1\%$ of the grid volume. This corresponds to shocks forming in cool intergranular lanes, where strong converging flows create supersonic compressions, also consistent with previous numerical studies (e.g., Fig.~6 of \citealt{steinSimulationsSolarGranulation1998}).

The CS filling factor (greens) remains approximately constant ($\sim50\%$ of the volume) from the subsurface of the CZ ($z=-0.5$~Mm) to $z=1$~Mm, before decreasing monotonically. This decline is linked to the decrease in plasma $\beta$ with height, increasing accordingly the magnetic field resistance to the flow shearing (see also Sect.~\ref{sec:currentS}). We also note two distinct peaks in the photosphere at $z=0$~Mm and $z=400$~km. The first one is due to the strong shearing of the magnetic field generated by convective flows, concentrating plasma into intergranular lanes. The second results from the overturning convection piling up horizontal field at the top of the photosphere \citep{steinerHorizontalInternetworkMagnetic2008}, which generates CS horizontally (see Fig.~\ref{fig:currentS_detection} and related animation).

Despite providing a global and average view of high-gradient process occurrences, the time- and spatially averaged filling factor is limited in understanding the subtle interplay of shocks and CS in chromospheric dynamics. To gain deeper insight into their direct impact on temperature structures, we will now analyze their evolution over the horizontal plane.

\subsection{Interplay with temperature structures in the chromosphere}\label{sec:interplay}
Fig.~\ref{fig:HorizLoc} illustrates temperature maps across three heights of interest: $z=0.5$~Mm, $z=1.0$~Mm and $z=-1.5$~Mm, for the left, middle, and right panel, respectively. We consider here a quarter of the horizontal surface to zoom in and clearly depict the dynamics. Shocks (purple) and CS locations (green) are overlaying grey-scaled temperature structures, where lighter regions correspond to hotter plasma.

Just before the chromosphere ($z=0.5$~Mm, left panel), shocks are not prominent yet, and CS (green) fill a significant part of the volume, as expected from Fig.~\ref{fig:fill_fact}. However, we do not report a clear correlation between CS and high-temperature regions (white areas), as the plasma $\beta>1$ regime limits the impact of CS dissipation via ohmic heating. Let us recall here that CS spread mainly horizontally at $z=500$~km as seen previously, and so horizontal cuts give rather thick overlays in the figure, although structures that we are tracking are notably thin.

At $z=1$~Mm (middle panel), the shocks become clearly visible (purple) and show a strong correlation with the high-temperature regions. Indeed, the Mach number $M=v/c_s\sim\sqrt{e_{\rm kin}/e_{\rm int}}$ exceeds 1 at these locations, meaning that the transfer term from the kinetic to the internal energy reservoir, i.e. via compression and viscous effects, can heat substantially if velocity gradients form. Shocks also strongly influence the dynamics, as the plasma $\beta$ is still on the order of unity or above, especially at shock locations. However, CS consequently do not show a strong correlation with high-temperature areas ($\beta\sim e_{\rm int}/e_{\rm mag}\gtrsim 1$). Indeed, some CS are observed overlying cooler regions (dark areas), indicating that CS-induced ohmic heating remains secondary to shock-driven heating dynamics at this depth.

Higher in the chromosphere ($z=1.5$~Mm, right panel), shocks continue to dominate the dynamics and to correlate with high-temperature regions. In addition, we now note that CS locations (green) now also clearly correlate with high-temperature structures (white areas). This is attributed to a lower plasma $\beta\lesssim1$ on average at this depth, where magnetic energy becomes dominant over internal energy. Thus, the former no longer limits the role of ohmic heating, whose strength correlates with CS location and can notably modify the local temperature structure. CS now further contribute to the thermal dynamics by altering the thermal balance as well at this height.

\section{Contribution to the heating}\label{sec:sect4}

\begin{figure*}
    \begin{center} 
        \includegraphics[width=0.49\linewidth]{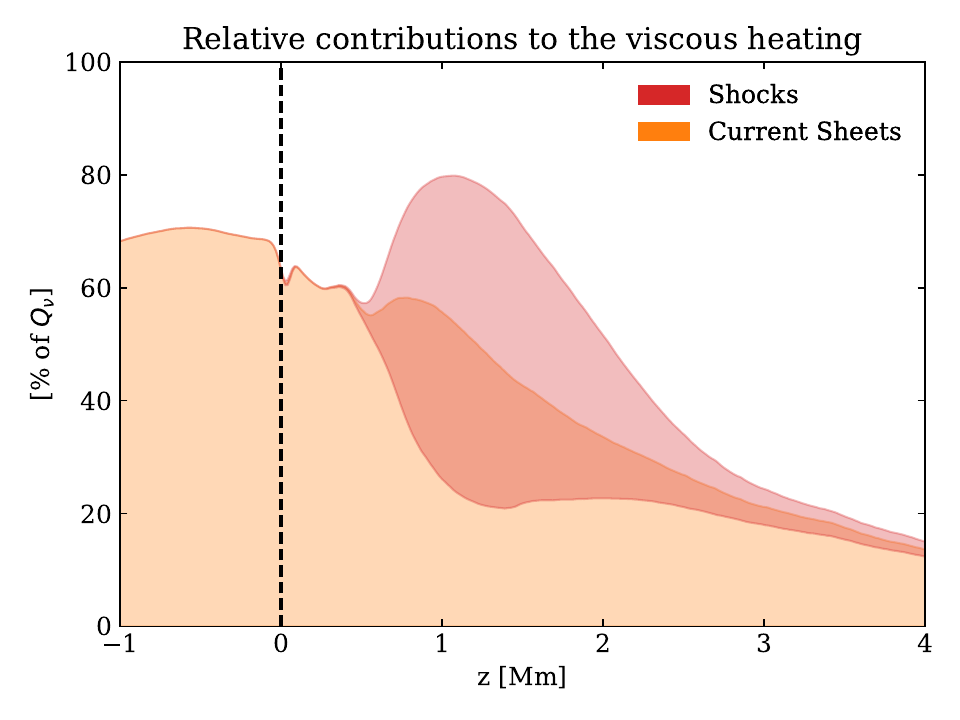}
        \includegraphics[width=0.49\linewidth]{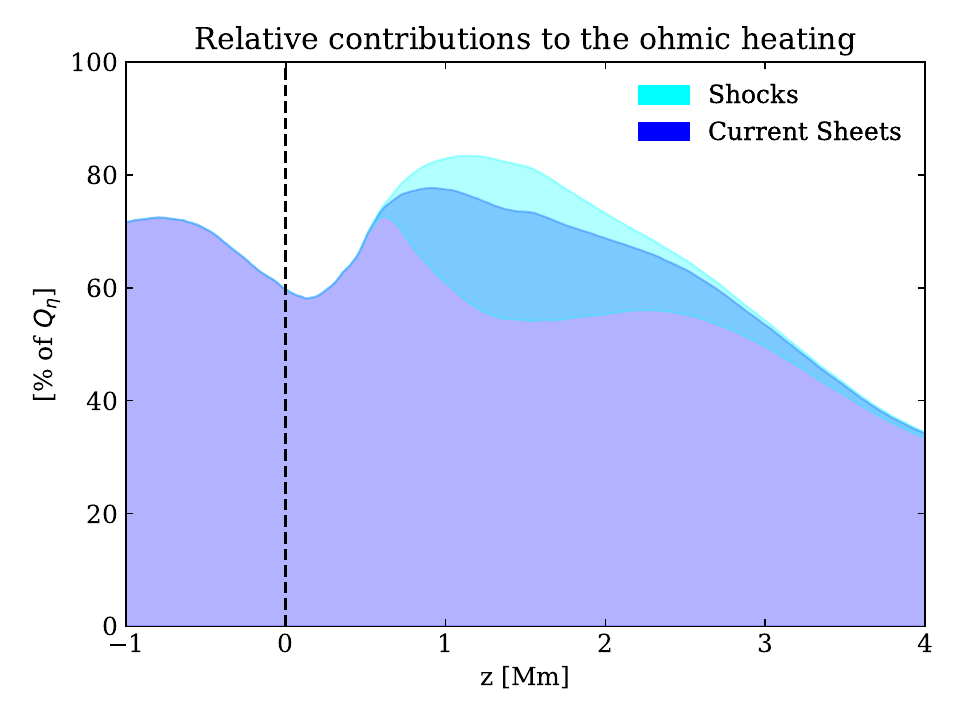}
    \end{center}
    \caption{Vertical profiles of the relative contributions of shocks and CS to heating via dissipation as a function of height. At each height, we sum either $Q_\nu$ (left) or $Q_\eta$ (right) over a selected subset of the domain (either shocks, CS, or the intersection of both) and compare it to the sum of the same quantity over the whole horizontal extent at that height. Profiles are averaged horizontally in space and over one solar hour in time. The vertical dashed line indicates the top of the CZ. \textit{Left}: Relative contribution to the viscous heating $Q_\nu$, with dark red for shocks and orange for CS. \textit{Right}: Relative contribution to the ohmic heating $Q_\eta$, with cyan blue for shocks and dark blue for CS.
    }\label{fig:StackVJ}
\end{figure*}

As mentioned in Sect.~\ref{sec:sect2}, the chromosphere marks the onset of the solar atmospheric heating. This acknowledges that part of the energy flux injected from the solar surface starts to be dissipated into heat in this layer. The chromospheric energy input can therefore be approximated from the total flux at its bottom, and the deposited energy from its decrease with height. We define
$$F_{\rm tot}=F_{\rm kin}+F_{en}+S$$
such as the sum of the kinetic, enthalpy and ideal Poynting fluxes, respectively. 

To set the lower boundary of the chromosphere, we adopt its canonical definition as the height where the photospheric radiative equilibrium approximation breaks \citep{schwarzschildEquilibriumSunsAtmosphere1906}. In practice, this occurs at $z = 600$~km in our model. Then, the upper boundary is set following the common usage that associates the chromosphere with the region of H$\alpha$ emission. We approximate this with $z = 2.57$~Mm, where the horizontally averaged temperature reaches 20~kK. Above this height, H$\alpha$ emission becomes negligible because hydrogen is mostly ionized \citep{carlssonApproximationsRadiativeCooling2012}.

We report an amplitude of total vertical flux $F_{\rm tot,z}(z_{\rm bot}=0.6~{\rm Mm})=5,2\times 10^6$~erg~cm$^{-2}$~s$^{-1}$ at the bottom of the chromosphere. This flux peaks at $5,7\times 10^6$~erg~cm$^{-2}$~s$^{-1}$ around $z=0.66$~Mm, before decreasing by two orders of magnitude down to $F_{\rm tot,z}(z_{\rm top}=2.57~{\rm Mm})=5,5\times 10^4$~erg~cm$^{-2}$~s$^{-1}$ at the top of the chromosphere. This chromospheric energy input estimate $F_{\rm tot,z}(z=0.6~{\rm Mm})$ is therefore consistent with chromospheric fluxes inferred for the quiet Sun from 1D semi-empirical models \citep{withbroeMassEnergyFlow1977,vernazzaStructureSolarChromosphere1981,andersonModelSolarChromosphere1989} and other 3D rMHD simulations \citep{martinez-sykoraOriginMagneticEnergy2019,tilipmanQuantifyingPoyntingFlux2023,khomenkoConvergenceStudyAmbipolar2025}.

In the following, we analyze how shocks and CS contribute to the heating that sustains the temperature structures we discussed. To this end, we focus on the \textit{mechanical} heating, which accounts for the local conversion of kinetic and magnetic energy into internal energy, via compression, viscous and ohmic heating. The analysis is carried out over one solar hour of simulation time between $t = 266$ and $t = 325$~minutes.

\subsection{Shocks and current-sheet contribution to the heating via viscous and ohmic dissipation}\label{sec:natHeatVJ}

As shocks and CS are fundamentally defined by strong gradients of the velocity and magnetic fields, respectively, they are natural sites of substantial energy dissipation, manifesting via viscous and ohmic heating. Consequently, one may directly associate viscous heating with shocks and ohmic heating with CS. However, Fig.~\ref{fig:StackVJ} clearly demonstrates that both structures contribute significantly to both forms of heating, highlighting the complex interplay between kinetic and magnetic processes in the chromospheric plasma.

Indeed, Fig.~\ref{fig:StackVJ} quantifies the relative contributions of shocks and CS to viscous (red, left panel) and ohmic (blue, right panel) heating as a function of height. The shock contribution partially overlaps with that of CS, as a given location can simultaneously satisfy both Eqs.~\ref{eq:cs_crit} and~\ref{eq:alpha_crit} criteria. This highlights the intertwined nature of these dissipation processes, particularly in the chromosphere, where the plasma $\beta=1$ surface lies. In this regime, shocks can initiate the formation of CS as they evolve, and conversely, CS may generate shocks through their collapse or magnetic reconnection.

Shocks exhibit peak contributions to viscous (dark red) and ohmic heating (cyan blue) at approximately $z = 1.2$ and $1.3\,\mathrm{Mm}$, respectively. These altitudes are close to the spatial maximum of shock occurrence (see Fig.~\ref{fig:fill_fact}), where approximately $8\%$ of the volume hence accounts for up to $57\%$ of the viscous heating and $28\%$ of the ohmic heating, illustrating the highly localized nature of energy dissipation driven in the \textit{Bifrost} simulation.

\begin{figure*}
    \begin{center}
        \includegraphics[width=0.49\linewidth]{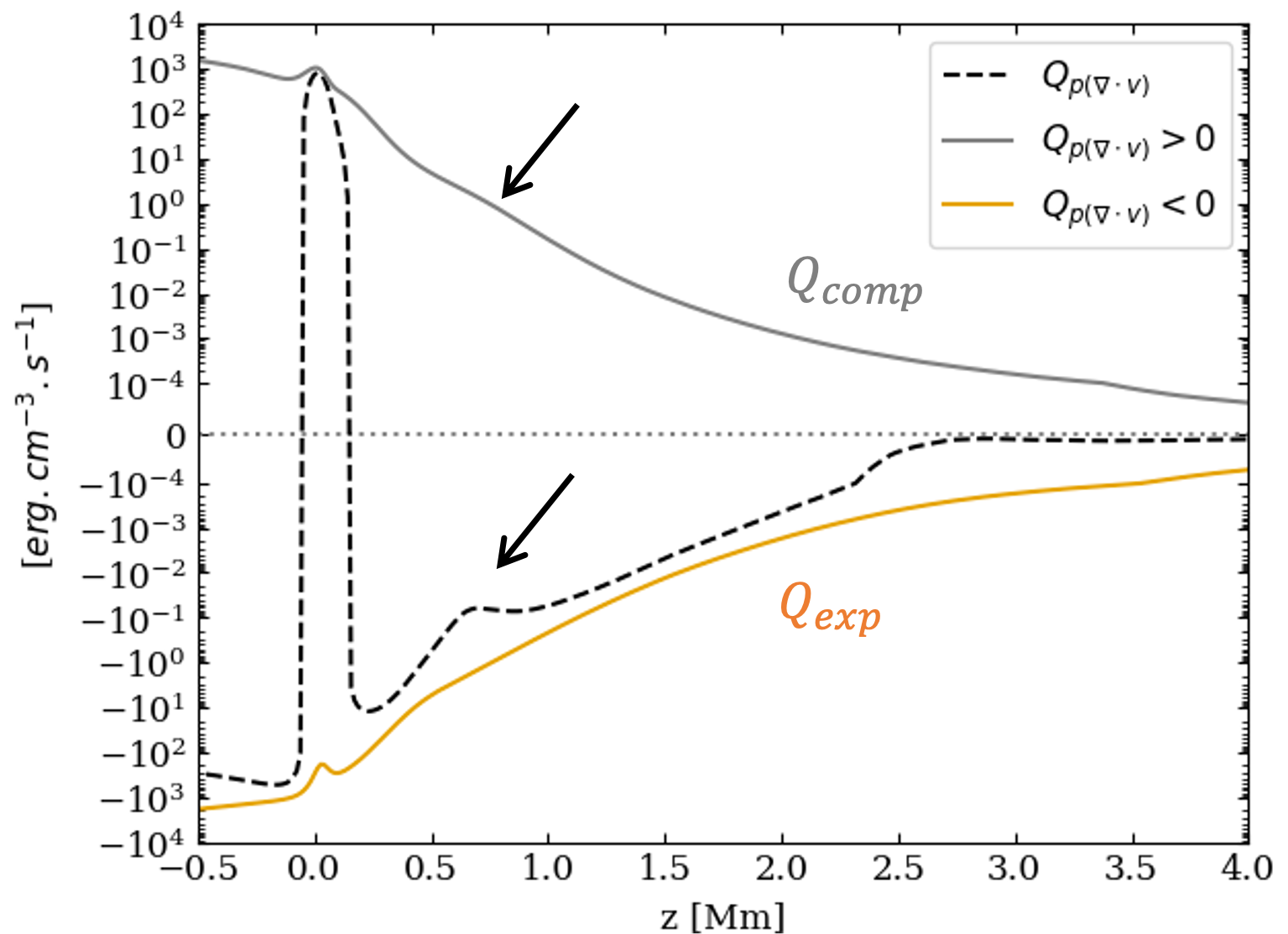}
        \includegraphics[width=0.49\linewidth]{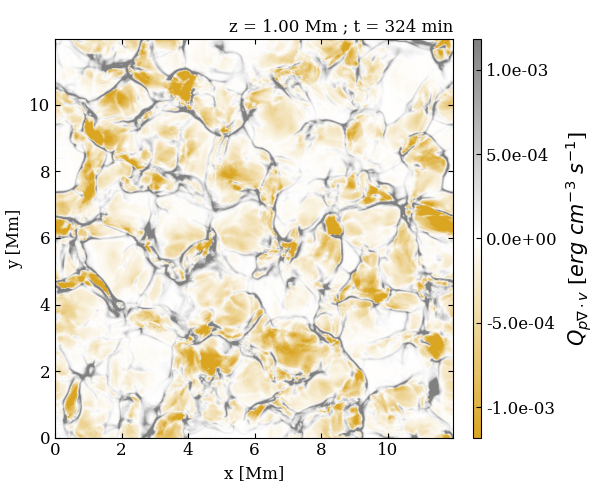}
    \end{center}
    \caption{Spatial distribution of the compressible term $Q_{p\nabla\cdot v}=-p\mathbf{\nabla}\cdot\mathbf{v}$ from the internal energy equation. \textit{Left}: The vertical profile is shown with a dashed black curve from the sub-surface CZ to the TR. Both positive (heating via compression $Q_{comp}$) and negative (cooling via expansion $Q_{exp}$) contributions of this term are illustrated with a grey and gold solid line, respectively. All profiles are averaged horizontally in space and over one solar hour in time. Dark arrows are pointing to the local increase of $Q_{p\nabla\cdot v}$, due to the change of behavior of $Q_{comp}$, resulting from upward-propagating waves that shock in the lower chromosphere. \textit{Right}: Horizontal cut of $Q_{p\nabla\cdot v}$ at $z=1$~Mm. Grey and gold regions illustrate heating via compression and cooling via expansion, respectively. The limits of the color bar correspond to twice the standard deviation of this map. An animation is available at (To be decided with AA) and covers 17~s, which corresponds to 27 solar minutes.}\label{fig:CompVsExp}
\end{figure*}

In the right panel of Fig.~\ref{fig:StackVJ}, CS are shown to dominate the total ohmic heating throughout the simulation domain. This is especially evident above the $\beta=1$ surface ($z \gtrsim 1\,\mathrm{Mm}$), where their contribution to ohmic heating remains significant despite a substantial decrease with height of their occurrence (see Fig.~\ref{fig:fill_fact}). For instance, more than $60\%$ of the ohmic heating arises from only $\sim 10\%$ of the volume around $z \sim 2\,\mathrm{Mm}$. This notable contribution, despite the application of a selective CS-detection criterion, further emphasizes the turbulent regime achieved in the chromosphere. The peak contribution of CS to ohmic heating reaches $78\%$ near $z \approx 0.9\,\mathrm{Mm}$, likely due to upward-propagating waves that shock, thereby forming small-scale magnetic gradients in the $\beta\gtrsim 1$ environment. This peak contribution from CS is indeed located at a lower altitude than that of the shocks ($z = 1.3$~Mm), as the average plasma $\beta$ decreases with height, progressively reducing the ability of shock-driven plasma motions to perturb and reconnect the magnetic field. This process, together with convective buffeting, may also promote interactions between the low-lying field and the larger loops or open field, particularly in the high-$\beta$ region.

Below $z = 0$, more than $60\%$ of both viscous and ohmic heating originates from regions identified as CS, despite an absence of shocks detected. This suggests that the dynamics in the CZ is predominantly driven by sub-sonic compression and shear flows, which subsequently create CS due to the high-$\beta$ regime (see Sect.~\ref{sec:sect3} for further details).

The remaining chromospheric heating contribution, neither due to shocks nor CS is resulting from broader gradients of $\mathbf{v}$ or/and $\mathbf{B}$, not considered as steep discontinuities by our criteria from Sect.~\ref{sec:sect3}. We suggest that this non-steep gradient fraction likely originates from weaker dissipation and compression mechanisms associated with propagating waves and shear-driven diffuse current layers, such as phase-mixing. Such features are similar to those indicated with arrows in Figs.~\ref{fig:shocks} and \ref{fig:currentS_detection}, though here located in the chromospheric vertical range. The detailed characterization of these structures in numerical simulations remains an active area of investigation and is left for future work.

\subsection{Heating via compression vs. cooling via expansion}\label{sec:CompVsExp}

Another major contributor to shock-induced and reconnecting-CS heating is the local compression (see e.g \citealt{wedemeyerNumericalSimulationThreedimensional2004,zafarEffectPlasma$v$Heating2025}), which is indeed essential for drawing a comprehensive picture of high-gradient processes in the chromospheric heating. We refer the interested reader to Appendix~\ref{sec:appC} for shock heating estimates via compression in an idealized chromosphere, and will here quantify it in the more complex chromosphere of our \textit{Bifrost} model.

Fig.~\ref{fig:CompVsExp} illustrates the spatial distribution of the compressible term $Q_{p\nabla\cdot v}$. This term represents the energy exchange between the kinetic and internal energy reservoirs, due to the work of pressure forces during local compression ($\nabla\cdot v<0$) and expansion ($\nabla\cdot v>0$). In the left panel, we illustrate its averaged profile with the dashed black line. As this term can be positive (heating through compression) or negative (cooling through expansion), we distinguish both contributions, denoted $Q_{\rm comp}$ (grey) and $Q_{\rm exp}$ (gold), respectively. The right panel further their horizontal distribution in a slice at $z = 1.00$~Mm.

\begin{figure*}
    \begin{center} 
        \includegraphics[width=0.49\linewidth]{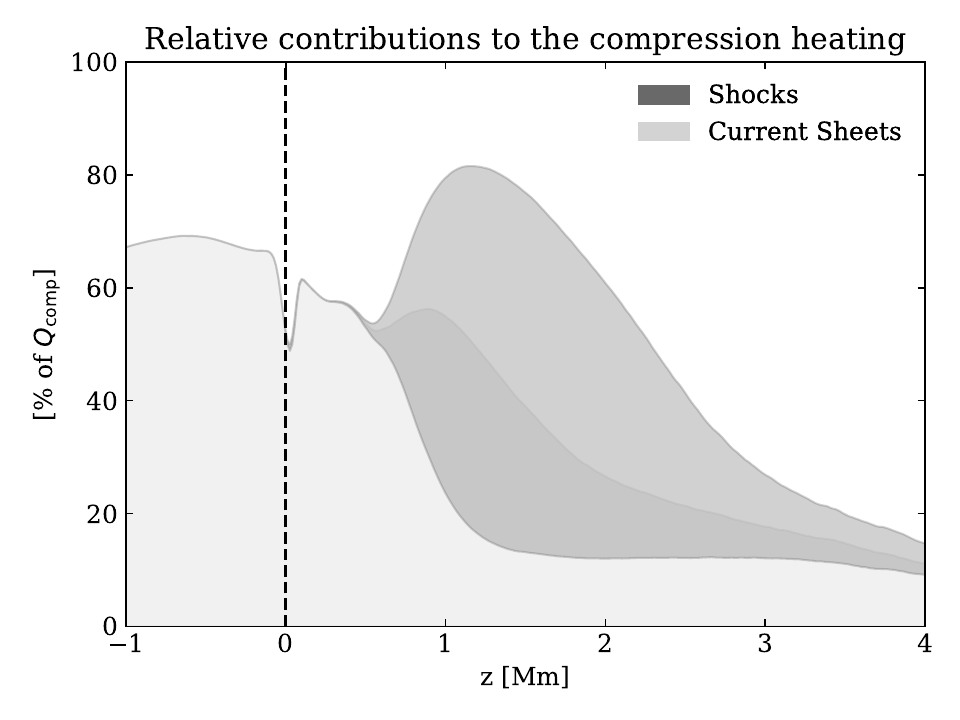}
        \includegraphics[width=0.49\linewidth]{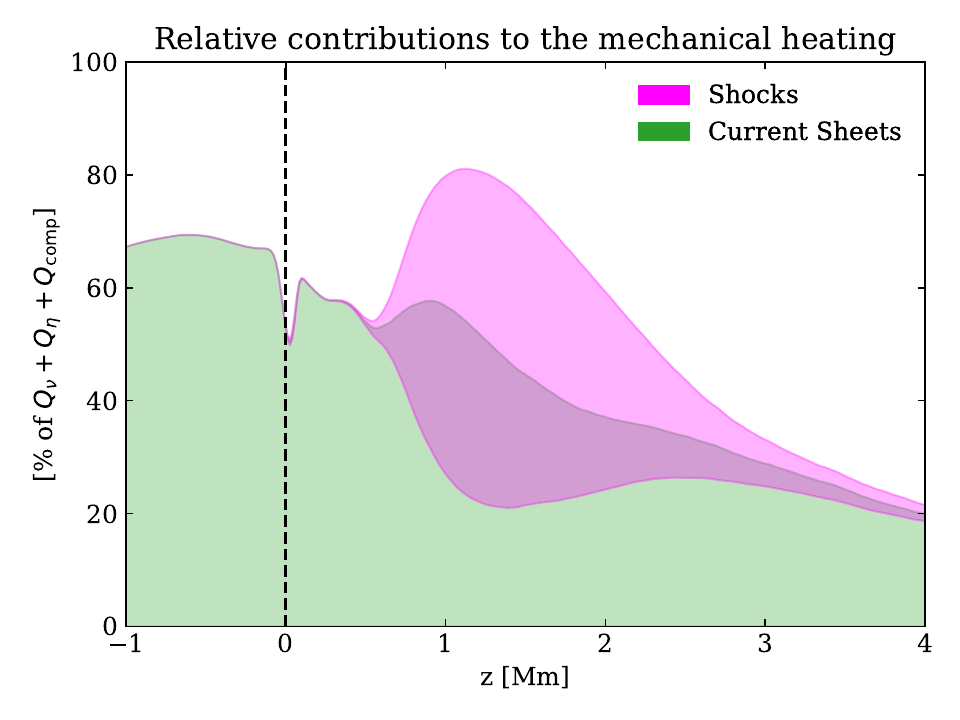}
    \end{center}
    \caption{Vertical profiles of the relative contributions of shocks and CS. Profiles are averaged horizontally in space and over one solar hour in time. At each height, we sum either $Q_{\rm comp}$ (left) or $Q_{\rm mech}=Q_\nu+Q_\eta+Q_{\rm comp}$ (right) over a selected subset of the domain (either shocks, CS, or the intersection of both) and compare it to the sum of the same quantity over the whole horizontal extent at that height. The vertical dashed line indicates the top of the CZ. \textit{Left:} Relative contribution to the compression heating $Q_{comp}$, with dark grey for shocks and light grey for CS. with the addition of dark grey for shocks and light grey for CS. \textit{Right}: Similar to the left panel in Fig.~\ref{fig:StackVJ}, but now adding up the compression heating from the left panel (see Appendix~\ref{sec:appD} for an illustration of the stacking, along with absolute values). Contributions of the shocks, CS, and non-steep gradients to the mechanical heating $Q_{\rm mech}$ are coloured in purple, green, and white, respectively.
    }\label{fig:StackVJC}
\end{figure*}

\begin{figure}
    \begin{center}
        \includegraphics[width=\linewidth]{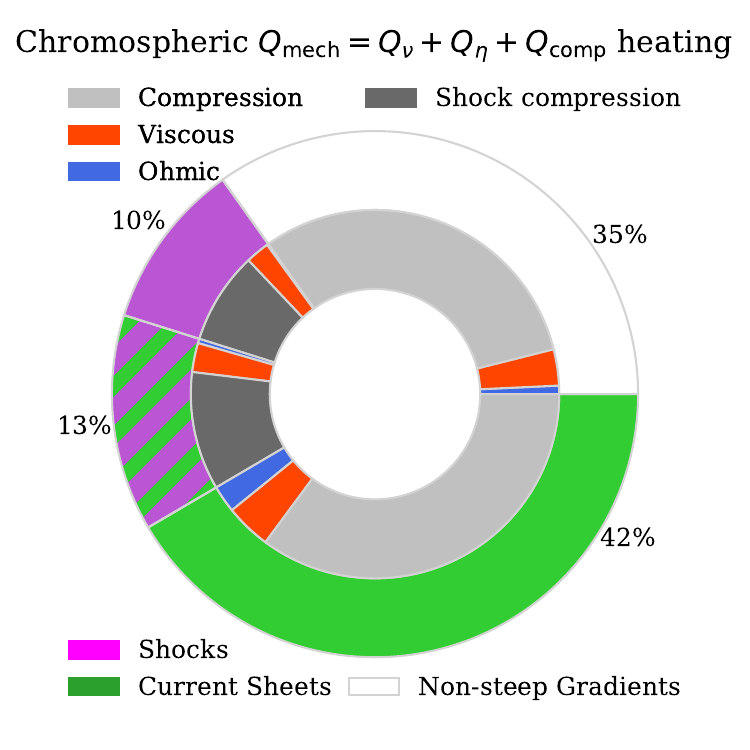}
    \end{center}
    \caption{Relative contributions of shocks (purple), CS (green) and non-steep gradients (white) to the mechanical heating of the chromosphere ($Q_{\rm mech}=Q_\nu + Q_\eta + Q_{comp}$, being red, blue and grey, respectively). The profiles used in the right panel of Fig.~\ref{fig:StackVJC} have been here spatially averaged over the chromospheric extent ($0.6\leq z\leq 2.57$~Mm), with the outer ring indicating the physical processes involved (shocks, CS, or neither), and the inner ring expliciting the associated dissipation mechanisms: viscous (red), ohmic (blue) and compression (grey). A darker shade of grey is used to highlight the shocks compression contribution (see also Appendix~\ref{sec:appA}). Hatched segments indicate the contribution of regions where both shocks and CS overlap.}\label{fig:PieVJC}
\end{figure}

First, we note that the compressible term $Q_{p\nabla\cdot v}$ is predominantly negative throughout most of the atmosphere, indicating that adiabatic expansion is the dominant process. This expansion arises primarily from granules overturning at the CZ surface, along with magnetically- and wave-driven expansion in the upper layers. The vertical profile also reveals a structured distribution with height. In particular, a clear local maximum of $Q_{p\nabla\cdot v}$ appears at $z=700$~km, associated with a change of behaviour in the compression contribution $Q_{comp}$ with height. This feature locates the onset of shock-induced compressional heating, which becomes significant at this altitude. The right panel further highlights how this heating contribution via shocks evolves in the horizontal plane (grey fronts), which we aim to quantify in the next section.

Two additional features stand out from the mean profile in the left panel. First, a local minimum at $z=230$~km corresponds to enhanced expansion when granules spread during their overturn. Second, a strong maximum near $z=0$~Mm is attributed to overshooting convective plumes that decelerate and compress when entering the convectively-stable layer above the photosphere. More generally, the dynamics of granulation involve granules pushing against each other and being disrupted by cool, descending material, leading to further compression and wave generation (see, e.g.,~\citealt{steinSimulationsSolarGranulation1998}). These waves steepen as they propagate upward and eventually form shocks, contributing to chromospheric heating.

\subsection{Respective contribution of processes to the mechanical heating}\label{sec:StackVJC}

Similarly to Sect.~\ref{sec:natHeatVJ}, we quantify and show the respective contributions of shocks (dark grey) and CS (light grey) to the compression heating $Q_{\mathrm{comp}}$, in the left panel of Fig.~\ref{fig:StackVJC}. Shock's contribution peaks at 66\% near $z = 1.3$~Mm. When combined with the CS contribution, the total power brought by these strong-gradient processes reaches 82\% at $z = 1.2$~Mm. The peak altitude of the CS contribution lies below that of shocks, around $z \sim 1$~Mm, again due to the average plasma $\beta$ decreasing with height (see also Sect.~\ref{sec:natHeatVJ}).

The right panel of Fig.~\ref{fig:StackVJC} then summarizes the relative contribution of shocks and CS to the atmospheric heating, when considering the \textit{mechanical heating} $Q_{\rm mech}$ as the sum of the viscous and ohmic dissipative heating with the one via compression, such that $Q_{\rm mech}=Q_\nu + Q_\eta + Q_{\mathrm{comp}}$. We here detail the vertical profile of the relative contributions, now focusing on processes rather than the nature of the energy deposition. This illustrates its different contributions by stacking again the one from compression (left panel) on top of those from viscous and ohmic dissipation (Fig.~\ref{fig:StackVJ}, see also Appendix~\ref{sec:appD}). This reveals that shocks and CS together can power up to 81\% of the mechanical heating $Q_{\rm mech}$ at $z = 1.1$~Mm. Notably, this significant fraction is concentrated in shocks/CS regions covering only 36\% of the horizontal area at that height. If we consider shocks only, their relative contribution reaches 59\% of the mechanical heating $Q_{\rm mech}$ at $z = 1.2$~Mm.

\subsection{Overall contributions to the chromospheric heating}\label{sec:PieVJC}

To obtain a global estimate for the chromosphere in Fig.~\ref{fig:PieVJC}, we vertically sum these contributions across its extent, defined here as $0.6 < z < 2.57$~Mm (see the beginning of Sect.~\ref{sec:sect4}).  Fig.~\ref{fig:PieVJC} then illustrates the overall chromospheric heating view in a pie chart, where the outer ring identifies the physical processes responsible for heating (shocks, CS, or neither), and the inner ring specifies the proportion in terms of deposition mechanism.

Shocks (purple) contribute for 23\%, and CS (green) 55\% of $Q_{\rm mech}$. When accounting for the overlap between shocks and CS, at least 66\% of the chromospheric heating can be attributed to these two processes. Given the conservative thresholds we applied for their identification (Sect.~\ref{sec:sect3}), we note here that these values can be seen as lower bounds of the shock/CS contributions.

By integrating the filling factors from Fig.~\ref{fig:fill_fact} over the chromospheric vertical extent, we find that shocks and CS occupy up to 6\% and 25\% of the volume, respectively. Accounting for their overlap (2\% of the volume), the combined volume fraction of shocks and CS is 29\%. Notably, shocks alone therefore contribute to 21\% of the total chromospheric mechanical heating while occupying only 6\% of the volume.

The inner ring of Fig.~\ref{fig:PieVJC} further details that compression (grey) dominates over viscous (red) and ohmic (blue) heating for all processes considered (shocks, CS, or neither). This is particularly pronounced for shocks, consistent with their genuine compressive nature. Interestingly, regions neither attributed to shocks nor CS still exhibit significant heating, primarily via compression as well. Given our conservative $\epsilon=6$ selection for steep gradients (Eq.~\ref{eq:cs_crit} and \ref{eq:alpha_crit}), part of this contribution may also include weak shocks that fall below detection thresholds. However, it appears this sensitivity is limited (see Appendix~\ref{sec:appB}), so this further reinforces the hypothesis that the non-steep gradients contribution also emerges from linear wave propagation, magnetic-field line braiding, and buffeting or emergence from the photosphere (see Figs.~\ref{fig:shocks} and \ref{fig:currentS_detection}). These aspects shall be further investigated in future studies (see, e.g.,~\citealt{udnaesCharacteristicsAcousticwaveHeating2025,cherryDetectionWaveActivity2025,enerhaugIdentifyingMagnetohydrodynamicWave2025}).

\section{Discussions}\label{sec:sect5}

\subsection{On the definition of atmospheric heating}\label{sec:defs}

Atmospheric heating, while fundamentally linked to a local increase in temperature, is subject to differences in definition depending on the context. Historically, the unexpected heating of the solar chromosphere and corona was identified via spectroscopic diagnostics (see Sect.~\ref{sec:sect1}). Consequently, observational studies often adopt a definition of \textit{heating} as any process that compensates for radiative losses.

From a theoretical and numerical perspective, however, heating can be more broadly understood as any process that contributes positively to the internal energy budget, given the proportionality between temperature and internal energy under solar conditions. For the sake of completeness and interpretation clarity, we choose to consider the \textit{mechanical heating} $Q_{\rm mech}=Q_\nu+Q_\eta+Q_{\rm comp}$, with $Q_{\rm comp}$ defined as the positive contributions of $Q_{\rm p\nabla\cdot v}$ (see Sect.~\ref{sec:CompVsExp}). This way, we can account for all processes that locally convert kinetic and magnetic energy into internal energy. 

It is important to recall here that the compression term $Q_{p\nabla\cdot v}$ can either act as a heating or cooling term, depending on the sign of local divergence. Some authors thus advocate for a more restrictive definition, which retains the irreversible viscous and ohmic components $Q_\nu$ and $Q_\eta$, which we here refer to as \textit{dissipative heating} (see Appendix~\ref{sec:appE}). However, excluding compression may underestimate significant heating, especially in chromospheric shock structures, where compression constitutes a leading-order heating mechanism (see, e.g., \citealt{carlssonNonLTERadiatingAcoustic1992,wedemeyerNumericalSimulationThreedimensional2004}). In chromospheric conditions, such compressive heating is unlikely to be fully reversible via adiabatic expansion. Instead, radiative cooling can rapidly drain the internal energy reservoir via non-local transport, effectively decoupling it from a subsequent and proportional feedback on the kinetic one locally. This asymmetry in energy exchange justifies the inclusion of heating via compression $Q_{\rm comp}$ in heating diagnostics, particularly for shocks.

Conversely, it should then also be recalled here that part of the internal energy created via the compression contribution (hence grey part in previous Figs.~\ref{fig:CompVsExp}, \ref{fig:StackVJC} and \ref{fig:PieVJC}) will not fully be drained via radiative emissions, but will also partially be transferred back via subsequent expansion (orange part in Fig.~\ref{fig:CompVsExp}). The exact ratio between the two possibilities is further modulated by local plasma parameters, such as density, ionization fraction, and magnetic topology (see, e.g., Chapter 6 of \citealt{mihalasFoundationsRadiationHydrodynamics1984}). These vary substantially with height in the chromosphere (see, e.g., Figs.~2 and 3 in \citealt{carlssonNonLTERadiatingAcoustic1992}) and a full quantification of the non-local transport \textit{vs.} reconversion terms would require a dedicated radiative transfer and MHD energy budget analysis, beyond the scope of the present study. 

For the sake of completeness and allowing comparisons, we thus present our results using both \textit{mechanical} and \textit{dissipative} definitions (inclusive and exclusive of $Q_{\rm comp}$, respectively, see Appendix~\ref{sec:appE}). Notably, the similarity in the ordering of relative heating contributions across both metrics (Figs.~\ref{fig:PieVJ} and \ref{fig:PieVJC}) shows that our conclusions are not dependent on the specific choice of definition, which reinforces the robustness of our heating identification and its relevance for solar atmospheric energetics.

\subsection{Solar context and caveats}\label{sec:caveats}

Our choice of strict detection criteria ($\epsilon=6$ in Eqs.~\ref{eq:cs_crit} and \ref{eq:alpha_crit}) ensures a minimal false-positive detection rate in gradient identification. Yet, it also means that we likely underestimate the total contributions of shocks and CS, meaning that the quantifications we give for both shocks and CS contributions should be regarded as lower bounds. Nevertheless, since our results exhibit limited sensitivity to the chosen value of the $\epsilon$ parameter (see Appendix~\ref{sec:appB}), we argue that the associated underestimation is similarly constrained.

The coronal temperatures in this experiment do not reach the 1~MK typically discussed in solar atmosphere simulations. Coronal temperatures are indeed dependent on magnetic topology within the computational domain and the amplitude of the average signed and unsigned B, which influence current formation and associated heating. Additionally, the domain size affects the capacity to channel energy from higher layers via thermal conduction (see, e.g., \citealt{carlssonPubliclyAvailableSimulation2016,finleyStirringBaseSolar2022} setups). In our case, the open magnetic configuration and relatively QS environment result in coronal temperatures (200 to 400~kK at heights of 5–8~Mm) compatible with those expected in quiet coronal holes \citep{cranmerCoronalHoles2009}. Nonetheless, the influence of magnetic configuration on the temperature structure remains a largely unconstrained parameter in numerical models and merits further investigation \citep{carlssonNewViewSolar2019}.

The model presented exhibits a box mode arising from the imposed pressure node at the bottom boundary. This mode is introduced intentionally to reproduce solar p-modes, although its spectrum is narrower because of the limited convection zone depth \citep{steinSolarOscillationsConvection2001}. As the mode period lies above the acoustic cut-off, related waves are evanescent, they thus do not steepen into shocks (see Fig. 6 of \citealt{carlssonPubliclyAvailableSimulation2016}). We therefore do not expect a substantial effect on the shock dynamics analyzed in this work.

To reduce computational cost, we did not consider a generalized Ohm's law. \cite{martinez-sykoraTWODIMENSIONALRADIATIVEMAGNETOHYDRODYNAMIC2012} demonstrated that ambipolar diffusion in the chromosphere is of a similar order of magnitude to the hyper-diffusion employed in Bifrost to ensure numerical stability. This supports the idea that the heating produced by the numerical diffusion scheme already has a realistic amplitude. However, this aspect inherently depends on the numerical scheme, which warrants for dedicated investigations in the future.

To reduce computational cost, we also neglected non-equilibrium ionization (NEI) effects. In the chromosphere, NEI prolongs ionization-recombination timescales, which are assumed instantaneous under LTE. For hydrogen and helium, these timescales exceed the characteristic ones of dynamic events such as magneto-acoustic shocks \citep{carlssonDynamicHydrogenIonization2002,leenaartsNonequilibriumHydrogenIonization2007}. As a result, the ionization state would lag behind rapid compressions or rarefactions, yielding smoother variations of the ionization fraction across shock fronts. This limits energy partitioning into ionization and enhances thermal heating, amplifying temperature fluctuations. Therefore, while NEI may affect peak temperature values (see, e.g., \citealt{martinez-sykoraIonNeutralInteractions2020}), we expect the impact on the total energy required to reach them, and thus on our conclusions, to be limited. This should be investigated in future work.

The Bifrost code employs a hyper-diffusivity scheme \citep{Nordlund1995,gudiksenStellarAtmosphereSimulation2011} that dynamically enhances resistivity in regions with sharp magnetic and velocity gradients. This ensures numerical stability during shocks and CS formation, while facilitating reconnection at realistic dissipation rates \citep{faerderComparativeStudyResistivity2023,faerderComparativeStudyResistivity2024,faerderExtremeultravioletEUVObservables2024}. It also means that the thickness of shocks and CS is set by the numerical diffusive scheme, typically six grid cells in Bifrost. In reality, such structures would indeed collapse to much thinner scales set by kinetic processes, which remain unresolved in current models (see, e.g., Chapter 5 of \citealt{mihalasFoundationsRadiationHydrodynamics1984} and \citealt{klimchukThicknessCurrentSheets2023}). As a result, the filling factors presented in Fig.~\ref{fig:fill_fact} are likely overestimated by comparison to a higher-resolution equivalent setup. However, we do not expect such a change to affect the net amount of energy dissipated by the detected structures, as the artificial diffusion scheme ensures resolution-independent energy deposition \citep{vonneumannMethodNumericalCalculation1950,Nordlund1995}, where related gradient amplitudes will accordingly increase. This was verified by reproducing the analysis on a shorter sample of a doubled-resolution simulation, yielding consistent results, with a quantification summarized in Appendix~\ref{sec:appB}. However, we expect that the time and spatial scales of the energy deposition may vary as the characteristic sizes and gradients evolve, an aspect that will be explored in future work.

\section{Conclusion and perspectives}\label{sec:sect6}

In this first part of our series on the QS chromosphere, we have performed a high-resolution 3D rMHD simulation using the \textit{Bifrost} code, and run a volumetric analysis of the energy dissipation. Shocks and CS arise naturally within the high-Reynolds environment of the chromosphere as sharp gradients of velocity and magnetic field, respectively, due to upwardly propagating acoustic waves, magnetic flux emergence, and twisting via magneto-convection. By applying physically motivated criteria to detect them, we characterized their spatial distribution, temporal evolution, and quantified their respective contributions to the chromospheric heating. The study was motivated by the long-standing need to identify deposition processes that are capable of compensating for the substantial chromospheric radiative losses observed on the Sun.

We emphasize here the necessity of including compression in the definition of heating, as discussed in Sect.~\ref{sec:defs}. In particular, the shocks produce significant local enhancements of internal energy through compression (darker grey parts in Fig.~\ref{fig:PieVJC}), which are neither fully reversible nor negligible in their thermodynamic impact. We also saw compression is a substantial heating contribution out of shocks (lighter grey parts in Fig.~\ref{fig:PieVJC}), both in CS and non-steep gradients. Excluding the compression contribution then risks overlooking critical heating, especially under chromospheric conditions where it has been reported as a cause of radiative cooling \citep{carlssonNonLTERadiatingAcoustic1992}. Accordingly, we here report our quantitative results in terms of the broader mechanical definition $Q_{\rm mech}=Q_\nu+Q_\eta+Q_{\rm comp}$, with $Q_{\rm comp}$ the positive compressive contribution. Nevertheless, interested readers can also find similar quantification in terms of the restrictive dissipative definition $Q_{\rm diss}=Q_\nu+Q_\eta$ (limited to viscous and Ohmic terms) in Appendix~\ref{sec:appE}.

The results indicate that shocks provide a dominant contribution to the energy deposition at lower chromospheric heights, as was already anticipated by earlier work \citep{carlssonNonLTERadiatingAcoustic1992,wedemeyer-bohmWhatHeatingQuietSun2007,kalkofenSolarChromosphereHeated2007,sobotkaCHROMOSPHERICHEATINGACOUSTIC2016,udnaesCharacteristicsAcousticwaveHeating2025}. The novelty of our results is to bring a clear quantification of how much energy is deposited by all shocks in a state-of-the-art 3D atmospheric model. This clarifies that shocks, while recurrent, contribute significantly only within a narrow vertical extent, here $1\lesssim z\lesssim 2$~Mm. Their contribution to the mechanical heating 59\%, at $z = 1.2$~Mm.

CS then provide the dominant contribution to chromospheric heating in the upper chromosphere ($1.5\lesssim z \lesssim 2.5$~Mm), due to
\begin{itemize}
    \item the increase of the local ohmic heating term in the internal energy balance due to the decrease of plasma-$\beta$ (see, e.g., Sect.~\ref{sec:interplay}),
    \item the occurrence decrease of shocks with height, as a substantial amount of their energy has already been deposited lower down, and as deflection due to wave conversion may take place when entering the lower-$\beta$ regime (see, e.g., \citealt{bogdanWavesMagnetizedSolar2003}).
\end{itemize} 
This provides a robust numerical support and quantification for the long-standing hypothesis that the dissipation of magnetic gradients and induced reconnection flows play a central role in the chromospheric energy budget \citep{beckEnergyWavesPhotosphere2009,jessMultiwavelengthStudiesMHD2015,carlssonNewViewSolar2019,abbasvandObservationalStudyChromospheric2020,molnarHighfrequencyWavePower2021}, especially in upper parts where shocks are "gone" \citep{steffensTracingCAGrains1997}. It is interesting to highlight that, even in CS-heating-dominated regions, viscous heating is of a similar order of magnitude to ohmic dissipation. This qualitatively confirms that velocity gradients are present in chromospheric CS and sustain here substantial viscous dissipation. This is consistent with the dynamics of reconnection outflows seen in Fig.~\ref{fig:currentS_detection}. Nevertheless, we must stress here that the exact value of the ratio between viscous and ohmic dissipation is sensitive to the numerical scheme and should be considered with care.

Once relative contributions are averaged over the full chromospheric extent (here $0.6< z< 2.57$~Mm), we report that shocks account for $23\%$ of the chromospheric $Q_{\rm mech}$, while CS account for $55\%$ of it. The union of both contributions represents $66\%$ of the mechanical chromospheric heating. Indeed, $13\%$ of the heating occurs in regions where both shocks and CS overlap. This co-spatiality directly results from the intricate physics occurring in the $\beta\sim 1$ chromospheric regime, and highlights the necessity of considering hybrid heating scenarios in chromospheric modeling, particularly for regions where observational proxies reveal ambiguous or multi-component signatures (see, e.g.,~\citealt{diazbasoObservationallyConstrainedModel2021,diaz-castilloConnectivitySolarPhotosphere2024}).

These results support the conclusion that small-scale strong-gradient intermittent events such as shocks or CS dominate the chromospheric heating under QS conditions, and quantify it in a state-of-the-art \textit{Bifrost} numerical model. However, it has been reported that the thermal structure may drastically change under different solar conditions \citep{vernazzaStructureSolarChromosphere1981,fontenlaEnergyBalanceSolar1993}, and more especially that the respective roles of each process are likely to be modified across different magnetic field configurations \citep{abbasvandChromosphericHeatingAcoustic2020,abbasvandObservationalStudyChromospheric2020}, which we aim to explore in future work.

In addition, Sect.~\ref{sec:shocks} has shown that, while a significant fraction of the upward acoustic flux is dissipated via shocks into chromospheric heating, a subset of it continues propagating when shocks undergo a transition back to linear wave behavior due to the steep increase of local sound speed in the TR. This sustains an injection of acoustic and Poynting flux into higher atmospheric layers, as evident in the animation of Fig.~\ref{fig:shocks}. The persistence of such features highlights the potential for coronal and solar-wind energization and warrants further investigation to assess their role in energy and momentum coupling across the chromosphere, TR, and low corona.

These results also reinforce the emerging picture of chromospheric heating as a fundamentally intermittent and multi-physics process, which is intrinsically challenging to capture without the use of comprehensive 3D rMHD models such as the one presented here. Such models could then provide key constraints on chromospheric heating for other types of numerical models commonly used in the community, such as 1-D field-aligned models (see, e.g.,~\citealt{carlssonNonLTERadiatingAcoustic1992}) as well as global models of the solar corona and wind, such as those employed in space weather applications, where chromospheric layers are not explicitly captured \citep{parsonsWangSheeleyArgeEnlilConeModel2011,vanderholstALFVENWAVESOLAR2014,pomoellEUHFORIAEuropeanHeliospheric2018,revilleRoleAlfvenWave2020,revilleFluxRopeDynamics2022,brchnelovaCOCONUTMFTwofluidIonneutral2023}.

Due to its small-scale and intermittent nature, shocks and CS dynamics are challenging to resolve observationally, particularly in QS regions where signal-to-noise ratios are low. The challenge becomes even more pronounced when attempting to resolve MHD wave processes operating in the linear regime, which is likely key to diagnosing the residual heating not driven by strong gradients. The present study then provides a quantitative benchmark and further guidance for upcoming observational efforts targeting the lower solar atmosphere. In this context, the emergence of next-generation facilities like the European Solar Telescope (EST, \citealt{quinteronodaEuropeanSolarTelescope2022}), will offer unprecedented spectral, spatial, and temporal resolution, critical for disentangling the intricate chromospheric physics.

These results support the view of chromospheric heating as an intermittent and multi-physics process that requires comprehensive 3D rMHD models to be properly resolved. Beyond quantifying the roles of shocks, and current sheets, such models can provide key constraints for widely used approaches, from 1-D field-aligned models (see, e.g.,~\citealt{carlssonNonLTERadiatingAcoustic1992}) to global coronal and solar wind simulations where the chromosphere is not explicitly captured \citep{parsonsWangSheeleyArgeEnlilConeModel2011,vanderholstALFVENWAVESOLAR2014,pomoellEUHFORIAEuropeanHeliospheric2018,revilleRoleAlfvenWave2020,brchnelovaCOCONUTMFTwofluidIonneutral2023}. At the same time, 3D models as the one presented here still need the support of observations to be better constrained. However, the small-scale and transient nature of shocks and CS remains difficult to resolve observationally, particularly in quiet-Sun regions, and even more for linear-wave dynamics. On this aspect, the higher resolution of next-generation facilities such as the European Solar Telescope (EST, \citealt{quinteronodaEuropeanSolarTelescope2022}) would further help disentangle this complex chromospheric dynamics.

\begin{acknowledgements}
The authors are thankful to M. Szydlarski, N. Poirier, G. Cherry, E. Enerhaug, F. Zang, J. Martinez Sikora, K. Krikova, E. R. Udnæs, and V. Hansteen for useful discussions. The authors also thank the anonymous referee for useful and constructive remarks. We acknowledge funding support by the European Research Council (ERC) under the European Union’s Horizon 2020 research and innovation programme (grant agreement No 810218 WHOLESUN), by the Research Council of Norway through its Centres of Excellence scheme (RoCS project number 262622), and through grants of computing time from the Programme for Supercomputing of Betzy/Sigma2, where the simulation was performed. The work of GA was supported by the Action Thématique Soleil-Terre (ATST) of CNRS/INSU PN Astro, also funded by CNES, CEA, and ONERA. Data manipulation was performed using the numpy \citep{harrisArrayProgrammingNumPy2020} and the in-house \textit{Bifrost} analysis pipeline \textit{helita} python packages. Figures in this work were produced using the python packages matplotlib \citep{HunterMatplotlib} and pyvista \citep{sullivan2019pyvista}.
\end{acknowledgements}

\bibliographystyle{aa}
\bibliography{MyLibrary}

%
%

\begin{appendix}

\section{Shocks detection calibration}\label{sec:appA}

\begin{figure}
    \begin{center}
        \includegraphics[width=\linewidth]{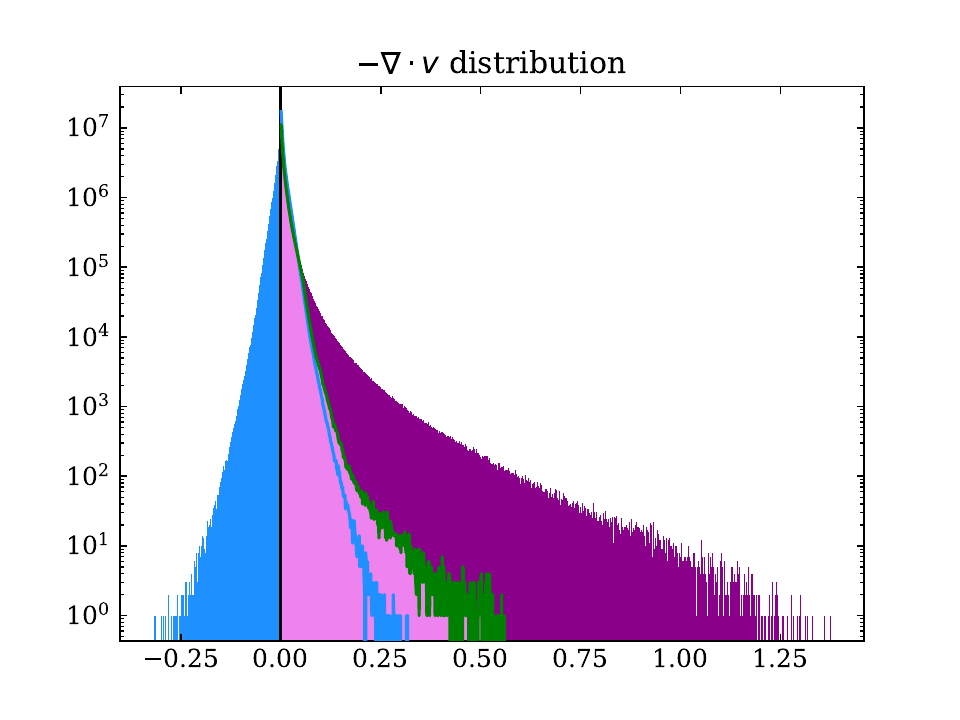}
    \end{center}
    \caption{Histogram of the compression frequency $-\mathbf{\nabla}\cdot\mathbf{v}$ in the whole domain. The negative (resp. positive) frequency population represents expansions (resp. compressions) and is colored in blue (resp. magenta). The green line represents the histogram of compression frequencies corresponding to $-\mathbf{\nabla}\cdot\mathbf{v}\leq\frac{c_s}{\epsilon.ds}$, with $\epsilon=6$ (see Eq.~\ref{eq:cs_crit})). We use a darker shade of magenta to highlight compressions attributed to non-linear shocks.}\label{fig:HistoShocks}
\end{figure}

In the linear regime, an acoustic wave propagating through a given location will tend to first compress and then expand the plasma in a quasi-reversible manner. The wave front is not steep enough to significantly diffuse kinetic or magnetic energy irreversibly into heat, so the wave tends to conserve its acoustic flux $F_{\rm acc}=\rho v^2 c_s$. However, if the velocity $v$ increases enough (due to e.g. a drop in sound speed $c_s$, density $rho$, or the intrisic tendency of a front to steepen, see, e.g., Burger's equation), it then becomes non-linear and turns into a shock, meaning that diffusion becomes significant enough to deposit part of the acoustic flux into heat irreversibly. One way to distinguish between linear and non-linear wave propagation is then to look at the $-\mathbf{\nabla.v}$ histogram in Fig.~\ref{fig:HistoShocks}. 

The compression-frequency distribution can be split into both its negative (expansion) and positive (compression) parts, colored in blue and magenta, respectively. In the positive part, we use the blue line to represent the symmetric of the negative (expansion) distribution for comparison. Linearly propagating waves are then supposed to lie in both, reversibly exchanging energy between the kinetic and internal energy reservoirs. Nevertheless, the positive tail (magenta) exhibits a non-symmetric high-frequency part due to front steepening. Shocks, which are inherently irreversible, will be located in the asymmetric part of the distribution. This non-linear population can then be selected using the sonic-compression criterion presented in Eq.~\ref{eq:cs_crit} (green line). We then use a darker shade of magenta to highlight compressions attributed to non-linear shocks.

The asymmetric part of the compression population, however, includes high-amplitude waves that one may not consider as shocks, depending on how strict the definition is, e.g., how thin the front has to be. In order to be selective, we want to select the thinnest front the code can resolve and one would unambiguously consider as a shock. The smallest typical scale emerging in a \textit{Bifrost} simulation is of the order of six grid points. We then choose to set the $\epsilon$ parameter of Eq.~\ref{eq:cs_crit} to 6 in this study, meaning that we aim to select wave fronts whose thickness is of the order of six grid points or less. The sensitivity of our results to this parameter is further explored in Appendix~\ref{sec:appB}.

\section{Sensitivity of the analysis to the $\epsilon$ parameter and resolution}\label{sec:appB}

\begin{table}
  \begin{center}
    \caption{Sensitivity analysis.}
    \begin{tabular}{lcccccc}
      \hline
      \hline
      $\epsilon$ & $Q_{\nu,sh}$ & $Q_{\nu,cs}$ & $Q_{\eta,sh}$ & $Q_{\eta,cs}$ & $Q_{{\rm comp},sh}$ & $Q_{{\rm comp},cs}$ \\
      (\#) & (\% $Q_{\rm mech}$) &  &  &  &  & \\[0.5ex]
      \hline
      \hline
      8  & 5.1 & 7.6 & 0.6 & 3.0 & 24.8 & 53.1 \\[0.5ex]
      7  & 4.9 & 7.1 & 0.6 & 2.9 & 21.7 & 59.7 \\[0.5ex]
      {\bf 6}  & {\bf 4.6} & {\bf 6.6} & {\bf 0.5} & {\bf 2.8} & {\bf 18.5} & {\bf 45.5} \\[0.5ex]
      5  & 4.3 & 5.9 & 0.4 & 2.6 & 15.1 & 40.5 \\[0.5ex]
      4  & 3.9 & 5.0 & 0.4 & 2.4 & 11.6 & 34.1 \\[0.5ex]
      \hline
      \hline
      {\bf 6}  & {\bf 5.3} & {\bf 5.4} & {\bf 0.4} & {\bf 2.4} & {\bf 18.5} & {\bf 38.6} \\[0.5ex]
      \hline
      \hline
    \end{tabular}
    \tablefoot{Shocks (\textit{sh}) and current-sheet (\textit{cs}) heating contributions integrated over $0.6\leq z\leq 2.57$~Mm (see Fig.~\ref{fig:PieVJC}). They are presented as fractions of the chromospheric mechanical heating $Q_{\rm mech}=Q_\nu+Q_\eta+Q_{\rm comp}$ and reported for different values of the $\epsilon$ parameter. The bottom line presents results for a similar run, where the resolution has been doubled.}\label{tab:sens}
  \end{center}
\end{table}

The identification of strong gradients requires choosing a value for the $\epsilon$ parameter, to define a threshold on the thickness of the front we want to consider (see Eqs.~\ref{eq:cs_crit} and \ref{eq:alpha_crit}). To test the sensitivity of our results to $\epsilon$, we report heating contributions for different values of it in Table~\ref{tab:sens}. Chromospheric contributions of shocks (\textit{sh}) and current sheets (\textit{cs}) to the viscous $Q_\nu$, ohmic $Q_\eta$ and compression heating $Q_{\rm comp}$ are given as percentages of the mechanical heating $Q_{\rm mech}$.

While the value of the parameter $\epsilon$ does impact the quantitative results, it does not alter the qualitative ordering among the different contributions. Both the \textit{sh} and \textit{cs} components increase with increasing $\epsilon$, as a broader gradient selection identifies a bigger number of fronts and sheets. Finally, we note that the sensitivity is slightly higher for the compressive heating term than for the dissipative ones, since the former scales with the first-order spatial derivative of the velocity field, while the latter depend on higher-order derivatives (second and third), which are more spatially localized.

Finally, we also present results for 20~minutes of an identical setup, where the resolution has been doubled, in the bottom line of Table~\ref{tab:sens}. When comparing both lines in bold, we note variations remain limited: the total shock contribution to $Q_{\rm mech}$ differs by $3\%$, while the total CS contribution differs by $18\%$., despite the factor 2 ($100\%$ change) on the resolution. Again, the overall impact does not alter the ordering of the contributions. Nevertheless, a dedicated study of the numerical convergence will be required to firmly conclude on the exact impact.

\begin{figure*}
    \begin{center} 
        \includegraphics[width=0.49\linewidth]{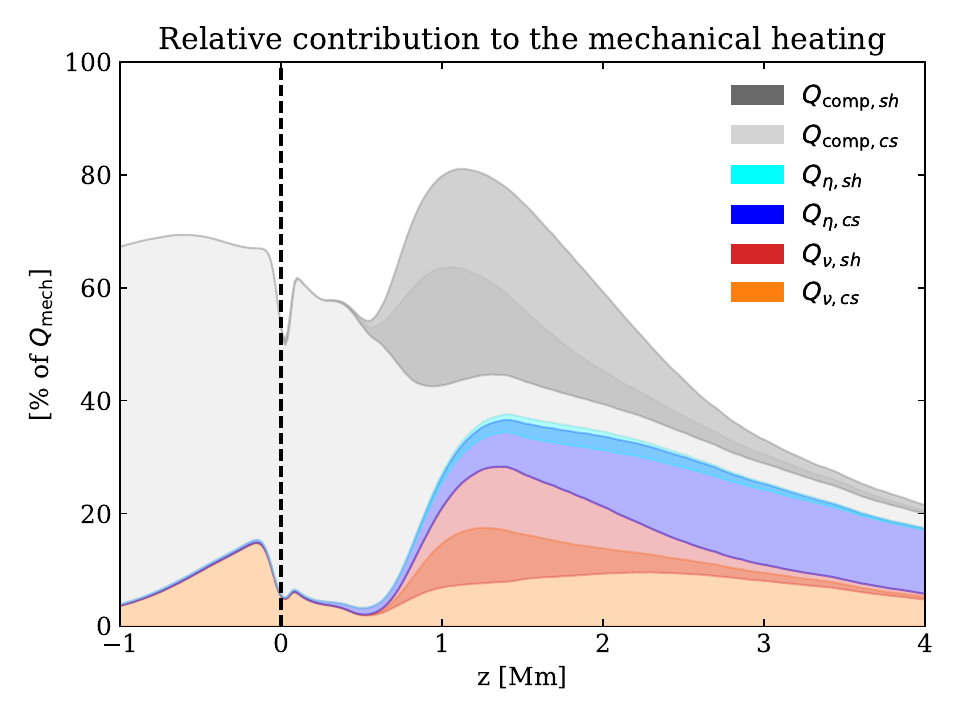}
        \includegraphics[width=0.49\linewidth]{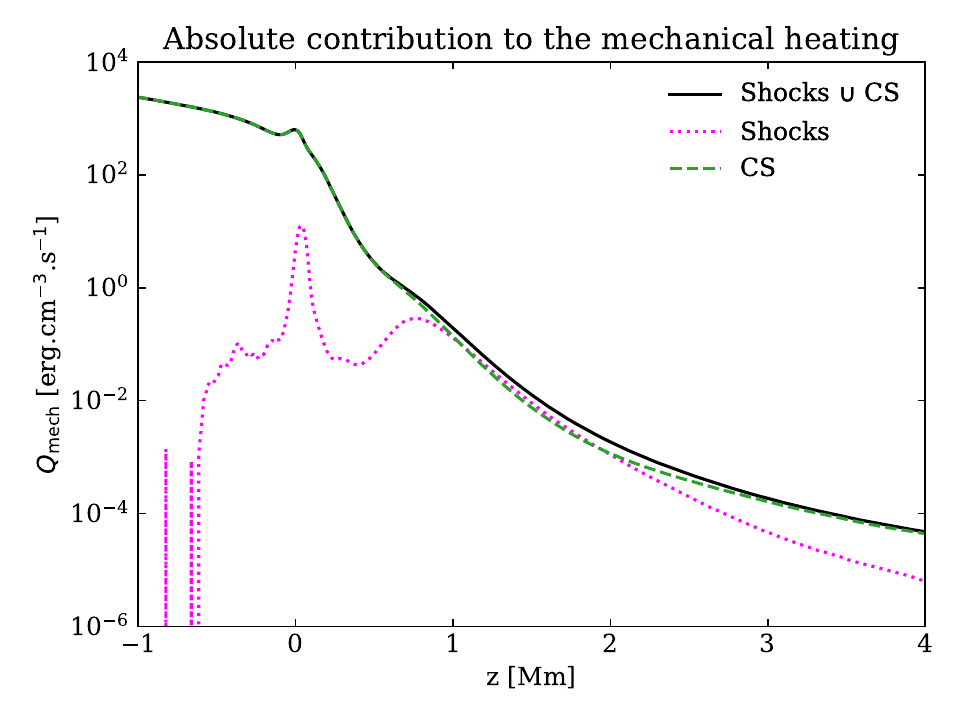}
    \end{center}
    \caption{\textit{Left}: Stack plot of the relative contributions of shocks, CS, and non-steep gradients (white), detailing the right panel of Fig.~\ref{fig:StackVJC}. The color code of the different relative contributions to the mechanical heating ($Q_{\rm mech}=Q_{\rm comp}+Q_\nu + Q_\eta$) is the same as for Figs.~\ref{fig:StackVJ} and \ref{fig:StackVJC}. Similarly, these contributions have been averaged in space horizontally and in time over one solar hour. \textit{Right}: Absolute contribution of shocks (dotted magenta), CS (dashed green), and the union of both (solid dark) to the mechanical heating.}\label{fig:StackVJCdetails}
\end{figure*}

\section{Idealized shock heating estimates}\label{sec:appC}

To complement the discussion about the importance of compression in shock-induced chromospheric heating, we recall the classical result for temperature jumps across a hydrodynamic shock, derived from Rankine-Hugoniot relations \citep{rankineXVThermodynamicTheory1870,ecole1887journal,ecole1889journal}. In the absence of magnetic fields ($B = 0$), the post-shock to pre-shock temperature ratio $T_2/T_1$ for an ideal gas with adiabatic index $\gamma$ and Mach number $M=v_1/c_{s,1}$ is given by
\begin{equation}
    \frac{T_2}{T_1} = \frac{(2\gamma M^2 - (\gamma - 1)) (2 + (\gamma - 1)M^2)}{(\gamma + 1)^2 M^2}.
\end{equation}

Assuming $\gamma = 5/3$ and an initial temperature of $T_1 = 6000$~K, the resulting post-shock temperatures are
\begin{align*}
    M = 1.5 &\Rightarrow T_2 \approx 1.49\, T_1 \approx 9\,000~\mathrm{K}, \\
    M = 2.0 &\Rightarrow T_2 \approx 2.08\, T_1 \approx 12\,000~\mathrm{K}, \\
    M = 3.0 &\Rightarrow T_2 \approx 3.67\, T_1 \approx 22\,000~\mathrm{K}.
\end{align*}

These estimates, here totally independent of viscosity and for Mach number values usually found in our simulation, illustrate that even mild supersonic compressions (e.g., $M \approx 1.5$) can induce heating to upper-chromospheric temperatures. Such idealized calculations offer a useful baseline for appreciating the key role of compression in shock-induced temperature fluctuations for chromospheric conditions.

\section{Absolute and details of the relative contributions to the mechanical heating.}\label{sec:appD}

To provide a broader view of the contributions of shocks and CS, we built the right panel of Fig.~\ref{fig:StackVJC} without distinguishing between the different heating mechanisms (compression, viscous, or ohmic). For clarity, these individual contributions are now detailed and illustrated in the left panel of Fig.~\ref{fig:StackVJCdetails}, which shows how they have been stacked. As discussed before, please note that some contributions overlap as some regions are classified as both shocks and CS, when they simultaneously satisfy the criteria of Eqs.~\ref{eq:cs_crit} and \ref{eq:alpha_crit}. Finally, the right panel in Fig.~\ref{fig:StackVJCdetails} presents the absolute values of the total contributions shown in Fig.~\ref{fig:StackVJC}.

\section{Chromospheric contribution to the dissipative heating}\label{sec:appE}

Reporting results for both $Q_{\rm mech}$ and $Q_{\rm diss}$ heating definitions ensures compatibility with earlier work, while preserving a consistent interpretation of the intricate nature of chromospheric heating (see Sect.~\ref{sec:defs}).

\begin{figure*}
    \begin{center} 
        \includegraphics[width=0.56\linewidth]{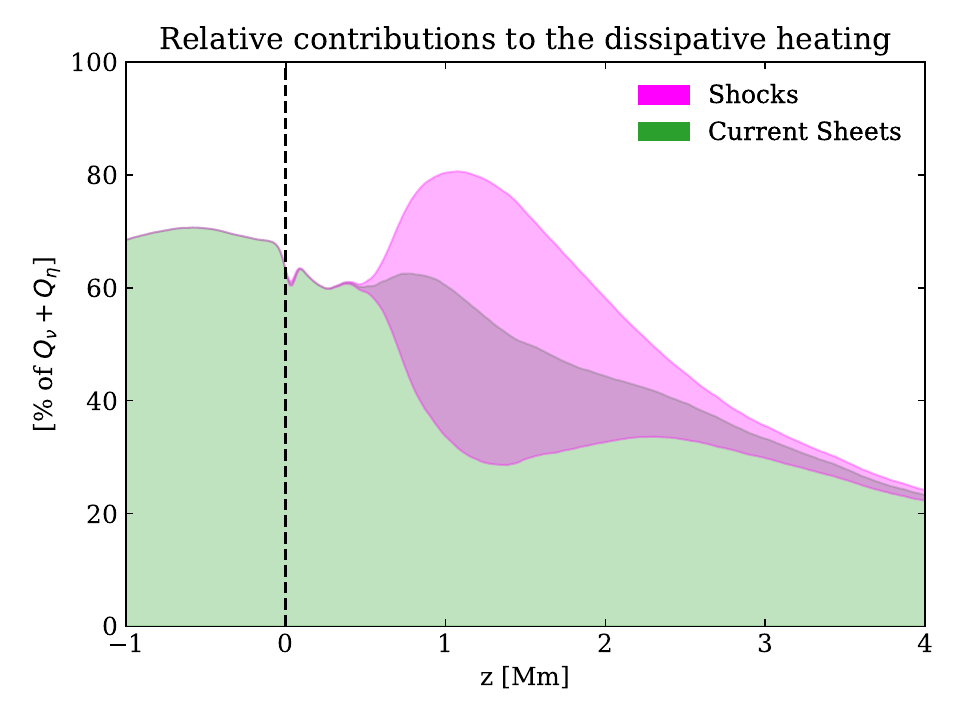}
        \includegraphics[width=0.42\linewidth]{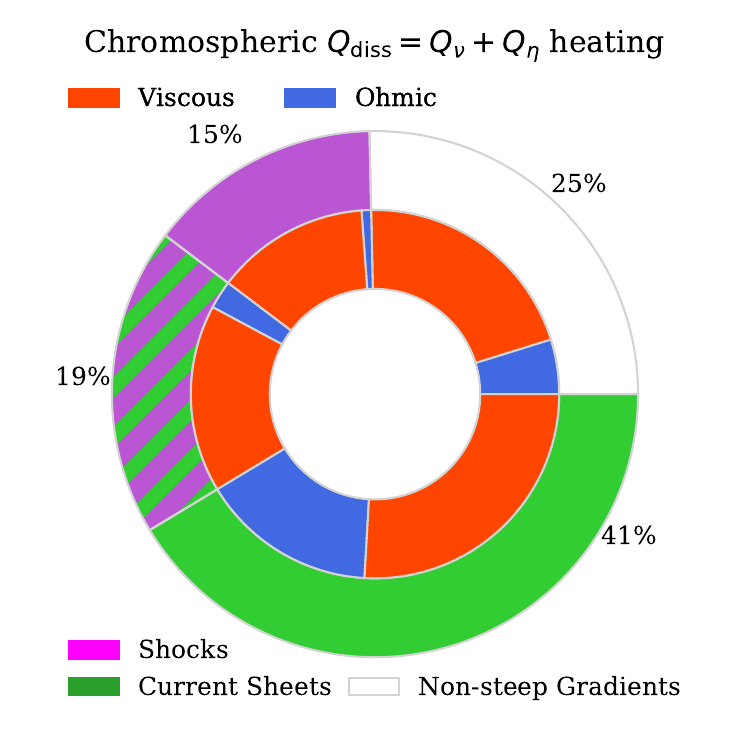}
    \end{center}
    \caption{Relative contributions of shocks (purple), CS (green) and non-steep gradients (white) to chromospheric dissipative heating ($Q_{\rm diss}=Q_\nu + Q_\eta$, red and blue respectively). \textit{Left:} Profile of the relative contributions to atmospheric dissipative heating as a function of height. At each height, we sum $Q_\nu+Q_\eta$ over a selected subset of the domain (either shocks, CS, or the intersection of both) and compare it to the sum over the whole horizontal extent at that height. Profiles are averaged horizontally in space and over one solar hour in time. The vertical dashed line indicates the top of the CZ.
    \textit{Right:} Same contributions, vertically averaged over the chromospheric extent ($0.6\leq z\leq 2.57$~Mm), with the outer ring indicating the physical processes involved (shocks, CS, or neither), and the inner ring expliciting the associated dissipation mechanisms (viscous vs. ohmic). Hatched segments indicate the contribution of regions where both shocks and CS overlap.}\label{fig:PieVJ}
\end{figure*}

We define here the \textit{dissipative heating} $Q_{\rm diss}$ as the sum of the viscous and ohmic heating $Q_{\rm diss}=Q_\nu + Q_\eta$. The right panel of Fig.~\ref{fig:PieVJ} then illustrates its different contributions by stacking the viscous and ohmic one, similarly to Sect.~\ref{sec:appD}. This yields the left panel, where we detail the vertical profile of such contributions, now focusing on processes (shocks and CS) rather than dissipation mechanisms (Fig.~\ref{fig:StackVJ}). We note that shocks primarily heat the lower chromosphere, with a 50\% peak contribution to $Q_{\rm diss}$ occurring close to $z=1.2$~Mm. When CS contributions are included, the cumulative amount reaches 81\% at $z=1.1$~Mm. The variation of the different contributions with height is consistent with our analysis in Sect.~\ref{sec:StackVJC}.

Now integrating over the chromosphere ($0.6 \leq z \leq 2.57$~Mm; see Sect.~\ref{sec:sect4}), we sum the contribution profiles from the left panel of Fig.~\ref{fig:PieVJ}, which yields the right panel. We see shocks (purple) account for 34\% and CS (green) for 60\% of the chromospheric dissipative heating $Q_{\rm diss}$. When accounting for their overlap, at least 75\% of the chromospheric dissipative heating can be attributed to such small-scale and strong-gradient events. This ordering of shocks and CS contributions are compatible with the one found with the mechanical heating definition in Sect.~\ref{sec:PieVJC}, showing our conclusions do not depend on the choice of heating definition.

The inner ring further reveals that viscous dissipation (red) dominates within shocks, as expected. For CS, viscous heating remains the primary mechanism of dissipation, despite the relative ohmic heating (blue) contribution increasing significantly. This suggests that related shears and reconnection flows play a key role in the dynamics and dissipation mechanisms of CS.

\end{appendix}

\section*{ORCID IDs}
Quentin NORAZ \orcidlink{0000-0002-7422-1127} \href{https://orcid.org/0000-0002-7422-1127}{https://orcid.org/0000-0002-7422-1127}\\
Mats CARLSSON \orcidlink{0000-0001-9218-3139} \href{https://orcid.org/0000-0001-9218-3139}{https://orcid.org/0000-0001-9218-3139}\\
Guillaume AULANIER \orcidlink{0000-0001-5810-1566} \href{https://orcid.org/0000-0001-5810-1566}{https://orcid.org/0000-0001-5810-1566}

\end{document}